\documentclass[11pt]{article}
\usepackage{times}
\usepackage{epsf}
\usepackage{psfig}
\newcommand{\cclass}[1]{{$\mathbf{#1}$}}

\newcommand{\newthmwithin}[3]{\newtheorem{#1q}{#2}[#3]
  \newenvironment{#1}{\begin{#1q}\sf}{\end{#1q}}}

\newcommand{\newthm}[3]{\newtheorem{#1q}[#2q]{#3}
  \newenvironment{#1}{\begin{#1q}\sf}{\end{#1q}}}

 \newthmwithin{theorem}{Theorem}{section}
\newthm{lemma}{theorem}{Lemma} \newthm{corollary}{theorem}{Corollary}
\newthm{definition}{theorem}{Definition} 
\newthm{fact}{theorem}{Fact}
\newthm{claim}{theorem}{Claim} \newthm{remark}{theorem}{Remark}
\newthm{observation}{theorem}{Observation}
\newthm{assumption}{theorem}{Assumption}

\newthm{problem}{theorem}{Problem}

\def\myendofproof{\vbox{\hrule\hbox{%
      \vrule height1.3ex\hskip0.8ex\vrule}\hrule}\par}

\newcommand{\set}[1]{\{ #1 \}}

\newcommand{\from}{from~}
\newcommand{\From}{From~}

\newcommand{\narrow}{\renewcommand{\baselinestretch}{.8}}
\newcommand{\wide}{\renewcommand{\baselinestretch}{1}}
\newcounter{program}
\newcounter{proglab}
\newcounter{proglabb}
\newcommand{\startstring}{}
\renewcommand{\theproglab}{\startstring\arabic{proglab}}
\renewcommand{\theproglabb}{\startstring\alph{proglabb}}

\newlength{\leftmarginsumii}
\newlength{\leftmarginsumiii}
\newlength{\leftmarginsumiv}
\newlength{\leftmarginsumv}
\newlength{\itemindentvalue} 
{\narrow\begin{list}{\theproglab\hfill}{\usecounter{proglab}
    \setlength{\topsep}{0pt} \setlength{\itemsep}{0pt}
    \setlength{\labelsep}{0pt} \setlength{\labelwidth}{\leftmargini}
    \setlength{\listparindent}{0pt}
    \setlength{\itemindent}{\itemindentvalue}
    \setlength{\leftmargin}{\leftmargini}} \setcounter{program}{0}
 \setcounter{proglab}{\value{program}}}%
{\end{list}\wide}

{\narrow\begin{list}{\theproglabb\hfill}{\usecounter{proglabb}
    \setlength{\topsep}{0pt} \setlength{\itemsep}{0pt}
    \setlength{\labelsep}{0pt} \setlength{\listparindent}{0pt}
    \setlength{\labelwidth}{\leftmargini}
    \setlength{\itemindent}{\itemindentvalue}
    \setlength{\leftmargin}{\leftmargini}} \setcounter{program}{0}
 \setcounter{proglabb}{\value{program}}}%
{\end{list}\wide}

{\narrow\begin{list}{\theproglab\hfill}{\usecounter{proglab}
    \setlength{\topsep}{0pt} \setlength{\itemsep}{0pt}
    \setlength{\labelsep}{0pt} \setlength{\listparindent}{0pt}
    \setlength{\labelwidth}{\leftmarginsumii}
    \setlength{\itemindent}{\itemindentvalue}
    \setlength{\leftmargin}{\leftmarginii}}
 \setcounter{proglab}{\value{program}}}%
{\end{list}\wide}

{\narrow\begin{list}{\theproglab\hfill}{\usecounter{proglab}
    \setlength{\topsep}{0pt} \setlength{\itemsep}{0pt}
    \setlength{\labelsep}{0pt} \setlength{\listparindent}{0pt}
    \setlength{\labelwidth}{\leftmarginsumiii}
    \setlength{\itemindent}{\itemindentvalue}
    \setlength{\leftmargin}{\leftmarginiii}}
 \setcounter{proglab}{\value{program}}}%
{\end{list}\wide}

{\narrow\begin{list}{\theproglab\hfill}{\usecounter{proglab}
    \setlength{\topsep}{0pt} \setlength{\itemsep}{0pt}
    \setlength{\listparindent}{0pt} \setlength{\labelsep}{0pt}
    \setlength{\labelwidth}{\leftmarginsumiv}
    \setlength{\itemindent}{\itemindentvalue}
    \setlength{\leftmargin}{\leftmarginiv}}
\setcounter{proglab}{\value{program}}}%
{\end{list}\wide}

{\narrow\begin{list}{\theproglab\hfill}{\usecounter{proglab}
    \setlength{\topsep}{0pt} \setlength{\itemsep}{0pt}
    \setlength{\listparindent}{0pt} \setlength{\labelsep}{0pt}
    \setlength{\labelwidth}{\leftmarginsumv}
    \setlength{\itemindent}{\itemindentvalue}
    \setlength{\leftmargin}{\leftmarginv}}
\setcounter{proglab}{\value{program}}}%
{\end{list}\wide}

 \newcommand{\calc}{{\cal C}}

\newcommand{\newspacing}{\baselineskip=1.2\normalbaselineskip}
\newcommand{\oldspacing}{\baselineskip= .9\normalbaselineskip}

\renewcommand{\thefootnote}{\fnsymbol{footnote}}
\input{epsf}

\begin{document}

\begin{titlepage}

  \title{ {\bf \Large Bicriteria Network Design Problems} }

\author{
{\sc Madhav V. Marathe }\footnotemark \and {\sc R. Ravi }\footnotemark
\and {\sc Ravi Sundaram }\footnotemark \and {\sc S. S. Ravi }\footnotemark
\and {\sc Daniel J. Rosenkrantz }\addtocounter{footnote}{-1}\footnotemark
\and {\sc Harry B. Hunt III }\addtocounter{footnote}{-1}\footnotemark
}
\maketitle

\begin{abstract}

\setcounter{footnote}{0} \addtocounter{footnote}{1} \footnotetext{Los
  Alamos National Laboratory, P.O. Box 1663, MS B265, Los Alamos, NM
  87545. Email: {\tt marathe@lanl.gov}.  Research supported by the
  Department of Energy under Contract W-7405-ENG-36.}
\addtocounter{footnote}{1} \footnotetext{GSIA, Carnegie Mellon
  University, Pittsburgh, PA 15213.  Email: {\tt ravi+@cmu.edu}. Work
  done at DIMACS, Princeton University, Princeton, NJ 08544. Research
  supported by a DIMACS postdoctoral fellowship and NSF Career Grant
   CCR-9625297.}
\addtocounter{footnote}{1} \footnotetext{
Delta Trading Co. Cambridge, MA. Work done while at 
LCS, MIT, Cambridge MA 02139.
  Email: {\tt koods@theory.lcs.mit.edu}. Research supported by DARPA
  contract N0014-92-J-1799 and NSF CCR 92-12184.}
\addtocounter{footnote}{1} \footnotetext{Department of Computer
  Science, University at Albany - SUNY, Albany, NY 12222.  Email:{\tt
    \{ravi, djr, hunt\}@cs.albany.edu}. Supported by NSF Grants CCR
  94-06611 and CCR 90-06396.} \addtocounter{footnote}{1}
\footnotetext{ A preliminary version of the the paper appeared in the
  {\em Proc. 22nd  International Colloquium on Automata
    Languages and Programming}, LNCS 944, pp. 487-498 (1995).}

We study a general class of bicriteria network design problems. A
generic problem in this class is as follows: Given an undirected graph
and two minimization objectives (under different cost functions), with
a budget specified on the first, find a <subgraph \from a given
subgraph-class that minimizes the second objective subject to the
budget on the first.  We consider three different criteria - the total
edge cost, the diameter and the maximum degree of the network.  Here,
we present the first polynomial-time approximation algorithms for a
large class of bicriteria network design problems for the above
mentioned criteria.  The following general types of results are presented.

First, we develop a framework for bicriteria problems and their
approximations.  Second, when the two criteria are the same 
we present a ``black box''
parametric search technique. This black box takes in as input an
(approximation) algorithm for the unicriterion situation and generates
an approximation algorithm for the bicriteria case with only a
constant factor loss in the performance guarantee. Third, when the two
criteria are the diameter and the total edge costs  
we use  a cluster-based approach to devise a 
approximation algorithms --- the solutions output violate both the
criteria by a logarithmic factor.
Finally, for the class of treewidth-bounded graphs, we provide
pseudopolynomial-time algorithms for a number of bicriteria problems
using dynamic programming. We show how these pseudopolynomial-time
algorithms can be converted to fully polynomial-time approximation
schemes using a scaling technique.

\end{abstract}


Network design, Combinatorial
algorithms.\\ 

\end{titlepage}

\newpage 

\newspacing

\renewcommand{\thefootnote}{\arabic{footnote}}

\section{Motivation}\label{sec:motivation}
With the information superhighway fast becoming a reality, the problem
of designing networks capable of accommodating multimedia (both audio
and video) traffic in a multicast (simultaneous transmission of data
to multiple destinations) environment has come to assume paramount
importance
\cite{Ch:Multicast,FW+:Multicast,KJ:Routing,KP+:Multicasting,KP+:Multicast}.
As discussed in  Kompella, Pasquale and Polyzos
\cite{KP+:Multicasting}, one of the
popular solutions to multicast routing involves tree construction.
Two optimization criteria -- (1) the minimum worst-case transmission
delay and (2) the minimum total cost -- are typically sought to
be minimized in the construction of these trees. Network design
problems where even one cost measure must be minimized, are often
\cclass{NP}-hard. 
(See Section A2 on Network Design in \cite{GJ:Computers}.)  
But, in real-life
applications, it is often the case that the network to be built is
required to minimize multiple cost measures simultaneously, with
different cost functions for each measure. For example, as pointed out
in \cite{KP+:Multicasting}, in the problem of finding good multicast
trees, each edge has associated with it two edge costs: the
construction cost and the delay cost. The construction cost is
typically a measure of the amount of buffer space or channel bandwidth
used and the delay cost is a combination of the propagation,
transmission and queuing delays.

Such multi-criteria network design problems, with separate cost
functions for each optimization criterion, also occur naturally in
Information Retrieval \cite{BK90} and 
VLSI designs (see \cite{ZP+:Iterative} and the references therein).
With the advent of deep micron VLSI designs, the feature size has
shrunk to sizes of 0.5 microns and less.  As a result, the
interconnect resistance, being proportional to the square of the
scaling factor, has increased significantly. An increase in
interconnect resistance has led to an increase in interconnect delays
thus making them a dominant factor in the timing analysis of VLSI
circuits.  Therefore VLSI circuit designers aim at finding minimum
cost (spanning or Steiner) trees given delay bound constraints on
source-sink connections.

The above applications set the stage for the formal definition of
multicriteria network design problems.  We explain this concept by giving
a formal definition of a bicriteria network design problem.
A generic bicriteria network
design problem, (\cclass{A}, \cclass{B}, \cclass{S}), is defined by
identifying two minimization objectives, - \cclass{A} and \cclass{B},
- \from a set of possible objectives, and specifying a membership
requirement in a class of subgraphs, - \cclass{S}.  The problem
specifies a budget value on the first objective, \cclass{A}, under one
cost function, and seeks to find a network having minimum possible
value for the second objective, \cclass{B}, under another cost
function, such that this network is within the budget on the first
objective. The solution network must belong to the subgraph-class
\cclass{S}.  For example, the problem of finding low-cost and
low-transmission-delay multimedia networks
\cite{KP+:Multicasting,KP+:Multicast} can be modeled as the (Diameter,
Total cost, Spanning tree)-bicriteria problem: given an undirected
graph $G = (V,E)$ with two  weight functions $c_e$ and $d_e$ for each
edge $e \in E$ modeling construction and delay costs respectively,
and a bound ${\cal D}$ (on the total delay), find a
minimum $c$-cost spanning tree such that the diameter of the tree
under the $d$-costs is at most ${\cal D}$. It is easy to see that the
notion of bicriteria optimization problems can be easily extended to
the more general multicriteria optimization problems.  In this paper,
we will be mainly concerned with bicriteria network design problems.

In the past, the problem of minimizing two cost measures was often
dealt with by attempting to minimize some combination of the two, thus
converting it into a unicriterion problem. This approach fails 
when the two criteria are very disparate. We have chosen,
instead, to model bicriteria problems as that of minimizing one
criterion subject to a budget on the other. We argue that this
approach is both general as well as robust. It is more general because
it subsumes the case where one wishes to minimize some functional
combination of the two criteria. It is more robust because the quality
of approximation is independent of which of the two criteria we impose
the budget on. We elaborate on this more in Sections \ref{sec:equivalence}
and \ref{sec:general}.

The organization of the rest of the paper is as follows: Section
\ref{sec:contributions} summarizes the results obtained in this paper;
Section \ref{sec:previous} discusses related research work; Section
\ref{sec:hardness} contains the hardness results; Section
\ref{sec:equivalence} shows that the two alternative ways of
formulating a given bicriteria problem are indeed equivalent; Section
\ref{sec:general} demonstrates the generality of the bicriteria
approach; Section \ref{sec:parametric} details the parametric search
technique; Section \ref{sec:diameter} presents the approximation
algorithm for diameter constrained Steiner trees; Section
\ref{sec:treewidth} contains the results on treewidth-bounded graphs;
Section \ref{sec:concluding} contains some concluding remarks and open
problems.

\section{Previous Work}\label{sec:previous}
\subsection{General Graphs}
The area of unicriterion optimization problems for network design is
vast and well-explored (See \cite{Ho:Approximation,CK:Compendium} and
the references therein.).  Ravi et al. \cite{RM+:Many} studied the
degree-bounded minimum cost spanning tree problem and provided an
approximation algorithm with performance guarantee ($O(\log n), O(\log
n)$). 


The (Degree, Diameter, Spanning tree) problem was studied by Ravi
\cite{Ra:Rapid} in the context of finding good broadcast networks.
There he provides an approximation algorithm for the (Degree, Diameter, 
Spanning tree) problem with performance guarantee ($O(\log^2 n),
O(\log n)$)\footnote{The result in Ravi \cite{Ra:Rapid} is actually
  somewhat stronger - given a budget, $D$, on the degree he finds a
  tree whose total cost is at most $O(\log n)$ times the optimal and
  whose degree is at most $O(D\log n + \log^2 n)$.}.

The (Diameter, Total cost, Spanning tree) entry in Table 1 corresponds
to the diameter-constrained minimum spanning tree problem introduced
earlier.  It is known that this problem is \cclass{NP}-hard even in
the special case where the two cost functions are identical
\cite{HL+:Bounded}. Awerbuch, Baratz and Peleg \cite{AB+:Cost} gave
an approximation algorithm with $(O(1),O(1))$ performance guarantee for
this problem - i.e. the problem of finding a spanning tree that has
simultaneously small diameter (i.e., shallow) and small total cost
(i.e., light), both under the same cost function. Khuller,
Raghavachari and Young \cite{KR+:Balancing} studied an extension
called {\em Light, approximate Shortest-path Trees (LAST)} and gave
an approximation algorithm with $(O(1),O(1))$ performance guarantee.
Kadaba and Jaffe
\cite{KJ:Routing}, Kompella et al.  \cite{KP+:Multicasting}, and Zhu
et al. \cite{ZP+:Iterative} considered the (Diameter, Total cost,
Steiner tree) problem with two edge costs and presented heuristics
without any guarantees. It is easy to construct examples to show that
the solutions produced by these heuristics in
\cite{ZP+:Iterative,KP+:Multicasting}, can be arbitrarily bad with
respect to an optimal solution.  
A closely related problem is that of
finding a diameter-constrained shortest path between two pre-specified
vertices $s$ and $t$, or (Diameter, Total cost, $s$-$t$ path). This
problem, termed the multi-objective shortest path problem (MOSP) in
the literature, is \cclass{NP}-complete and Warburton
\cite{Wa:Approximation} presented the first fully polynomial
approximation scheme (\cclass{FPAS}) for it. Hassin
\cite{Ha:Approximation} provided a strongly polynomial \cclass{FPAS}
for the problem which improved the running time of Warburton
\cite{Wa:Approximation}. This result was further improved by 
Phillips \cite{Ph:Network}.

The (Total cost, Total cost, Spanning tree)-bicriteria problem has
been recently studied by Ganley et al. \cite{GG+:Multi-weighted}.
They consider a more general problem with more than two weight
functions.  They also gave approximation algorithms for the restricted
case when each weight function obeys triangle inequality. However,
their algorithm does not have a bounded performance guarantee with
respect to each objective.

\subsection{Treewidth-Bounded Graphs}

Many \cclass{NP}-hard  problems have exact solutions when
attention is restricted to the class of treewidth-bounded graphs and
much work has been done in this area (see
\cite{AC+:Algebraic,AL+:Easy,BL+:Linear} and the references therein).
Independently, Bern, Lawler and Wong \cite{BL+:Linear} introduced the
notion of decomposable graphs. Later, it was shown
\cite{AC+:Algebraic} that the class of decomposable graphs and the
class of treewidth-bounded graphs are equivalent. Bicriteria network
design problems restricted to treewidth-bounded graphs have been
previously studied in \cite{AL+:Easy,Bo:Dynamic}.


\section{Our Contributions}\label{sec:contributions}
In this paper, we study the complexity and approximability of a number
of bicriteria network design problems.  
The three objectives we
consider are: (i) total cost, (ii) diameter and (iii) degree of the
network.  These reflect the price of synthesizing the network, the
maximum delay between two points in the network and the reliability of
the network, respectively. The {\em Total cost} objective is the sum
of the costs of all the edges in the subgraph.  The {\em Diameter}
objective is the maximum distance between any pair of nodes in the
subgraph.  The {\em Degree} objective denotes the maximum over all nodes
in the subgraph, of the degree of the node.  
The class of subgraphs we
consider in this paper are mainly {\em Steiner trees} (and hence {\em
  Spanning trees} as a special case); although several of our results
extend to more general connected subgraphs such as generalized Steiner
trees.

As mentioned in \cite{GJ:Computers}, most of the problems considered
in this paper, are \cclass{NP}-hard for arbitrary instances even when
we wish to find optimum solutions with respect to a single criterion.
Given the hardness of finding optimal solutions, we concentrate on
devising approximation algorithms with worst case performance
guarantees.  Recall that an approximation algorithm for a
minimization problem $\Pi$ provides a {\bf performance guarantee} of
$\rho$ if for every instance $I$ of $\Pi$, the solution value returned
by the approximation algorithm is within a factor $\rho$ of the
optimal value for $I$.  Here, we extend this notion to apply to
bicriteria optimization problems.  An $(\alpha,\beta)$-approximation
algorithm for an (\cclass{A}, \cclass{B}, \cclass{S})-bicriteria problem
is defined as a polynomial-time algorithm that produces a solution in
which the first objective (\cclass{A}) value, is at most $\alpha$
times the budget, and the second objective (\cclass{B}) value, is at
most $\beta$ times the minimum for any solution that is within the
budget on \cclass{A}. The solution produced must belong to the
subgraph-class \cclass{S}. Analogous definitions can be given when 
\cclass{A} and/or \cclass{B} are maximization objectives. 

\subsection{General Graphs}
Table 1 contains the performance guarantees of our approximation
algorithms for finding spanning trees, \cclass{S}, under different
pairs of minimization objectives, \cclass{A} and \cclass{B}.  For each
problem cataloged in the table, two different costs are specified on
the edges of the undirected graph: the first objective is computed
using the first cost function and the second objective, using the
second cost function.  The rows are indexed by the budgeted objective.
For example the entry in row \cclass{A}, column \cclass{B}, denotes
the performance guarantee for the problem of minimizing objective
\cclass{B} with a budget on the objective \cclass{A}.  All the results
in Table 1 extend to finding Steiner trees with at most a constant
factor worsening in the performance ratios.  For the diagonal entries
in the table the extension to Steiner trees follows \from Theorem
\ref{better-scale-thm}. {\sc Algorithm DCST} of Section
\ref{sec:diameter} in conjunction with {\sc Algorithm
  Bicriteria-Equivalence} of Section \ref{sec:equivalence} yields the
(Diameter, Total cost, Steiner tree) and (Total cost, Diameter,
Steiner tree) entries.  The other nondiagonal entries can also be
extended to Steiner trees and these extensions will appear in the
journal versions of \cite{RM+:Many,Ra:Rapid}.  Our results for
arbitrary graphs can be divided into three general categories.

{\footnotesize
\begin{center}\makebox[0in]{
\begin{tabular}{|c|c|c|c|}
  \hline

Cost Measures  &  Degree    &  Diameter   &  Total Cost\\ \hline
Degree         &   $(O(\log n), O(\log n)) ^*$ &  $(O(\log^2 n), O(\log n))$\cite{Ra:Rapid} & $(O(\log n), O(\log n))$\cite{RM+:Many}\\ \hline 
Diameter       &   $(O(\log n), O(\log^2 n))$\cite{Ra:Rapid}      &   $(1 +
\gamma, 1 + \frac{1}{\gamma})^*$    & 
                                            $(O(\log n), O(\log n))^*$\\ \hline 
Total Cost    & $(O(\log n), O(\log n))$\cite{RM+:Many} &  $(O(\log n),O(\log n))^*$ & 
                                                        $(1 +
\gamma, 1 + \frac{1}{\gamma})^*$\\ \hline 
\end{tabular}
}

\vspace*{.1in}

{\bf Table 1. Performance Guarantees for finding spanning trees in an
arbitrary graph on $n$ nodes. Asterisks indicate results obtained in
this paper. $\gamma > 0$ is a fixed accuracy parameter.}\\
\end{center}
} 

First, as mentioned before, there are two natural alternative ways of
formulating general bicriteria problems: (i) where we impose the
budget on the first objective and seek to minimize the second and (ii)
where we impose the budget on the second objective and seek to
minimize the first. We show that an $(\alpha,\beta)$-approximation
algorithm for one of these formulations naturally leads to a
$(\beta,\alpha)$-approximation algorithm for the other.  Thus our
definition of a bicriteria approximation is independent of the choice
of the criterion that is budgeted in the formulation. This makes it a
robust definition and allows us to fill in the entries for the
problems (\cclass{B}, \cclass{A}, \cclass{S}) by transforming the
results for the corresponding problems (\cclass{A}, \cclass{B},
\cclass{S}).

Second, the diagonal entries in the table follow as a corollary of a general
result (Theorem \ref{better-scale-thm}) which is proved using a
parametric search algorithm.  The entry for (Degree, Degree, Spanning
tree) follows by combining Theorem \ref{better-scale-thm} with the
$O(\log n)$-approximation algorithm for the degree problem in
\cite{RM+:Many}. In \cite{RM+:Many} they actually provide an $O(\log
n)$-approximation algorithm for the weighted degree problem. The
weighted degree of a subgraph is defined as the maximum over all nodes
of the sum of the costs of the edges incident on the node in the
subgraph. Hence we actually get an $(O(\log n), O(\log
n))$-approximation algorithm for the (Weighted degree, Weighted
degree, Spanning tree)-bicriteria problem. Similarly, the entry for
(Diameter, Diameter, Spanning tree) follows by combining Theorem
\ref{better-scale-thm} with the known exact algorithms for minimum
diameter spanning trees \cite{CG:Bounded}; while the
result for (Total cost, Total cost, Spanning tree) follows by
combining Theorem \ref{better-scale-thm} with an exact algorithm to
compute a minimum spanning tree \cite{CLR}.

Finally, we present a cluster based approximation algorithm and a
solution based decomposition technique for devising approximation
algorithms for problems when the two objectives are different. Our
techniques yield $(O(\log n), O(\log n))$-approximation algorithms for
the (Diameter, Total cost, Steiner tree) and the (Degree, Total cost,
Steiner tree) problems\footnote{The result for (Degree, Total cost,
  Steiner tree) can also be obtained as a corollary of the results in
  \cite{RM+:Many}.}.

\subsection{Treewidth-Bounded Graphs}

We also study the bicriteria problems mentioned above for the class of
treewidth-bounded graphs. Examples of treewidth-bounded graphs include
trees, series-parallel graphs, $k$-outerplanar graphs, chordal graphs
with cliques of size at most $k$, bounded-bandwidth graphs etc.  We
use a dynamic programming technique to show that for the class of
treewidth-bounded graphs, there are either polynomial-time or
pseudopolynomial-time algorithms (when the problem is {\bf
  NP}-complete) for several of the bicriteria network design problems
studied here.  A {\bf polynomial time approximation
scheme} ({\sf PTAS}) for problem $\Pi$ is a family of algorithms
${\cal A}$
such that, given an instance $I$ of $\Pi$, 
for all $\epsilon > 0$, there
is a polynomial time algorithm $A \in {\cal A}$  that returns a solution
which is within a factor $(1+\epsilon)$ of the optimal value for $I$.
A polynomial time approximation scheme in which the running time grows
as a polynomial function of $\epsilon$ is called a {\bf fully
  polynomial time approximation scheme}.  Here we show how to convert
these pseudopolynomial-time algorithms for problems restricted to
treewidth-bounded graphs into fully polynomial-time approximation
schemes using a general scaling technique.  Stated in our notation, we
obtain polynomial time approximation algorithms with performance of
$(1, 1+ \epsilon)$, for all $\epsilon > 0$.  The results for
treewidth-bounded graphs are summarized in Table 2.  As before, the
rows are indexed by the budgeted objective. All algorithmic results in
Table 2 also extend to Steiner trees in a straightforward way.

Our results for treewidth-bounded graphs have an interesting
application in the context of finding optimum broadcast schemes.
Kortsarz and Peleg \cite{KP:Approximation} gave $O(\log
n)$-approximation algorithms for the minimum broadcast time problem
for series-parallel graphs. Combining our results for the (Degree,
Diameter, Spanning tree) for treewidth-bounded graphs with the
techniques in \cite{Ra:Rapid}, we obtain an 
$O(\frac{ \log n}{\log \log n})$-approximation algorithm for
the minimum broadcast time  problem for 
treewidth-bounded graphs (series-parallel graphs have a treewidth of
$2$), improving and generalizing the result in \cite{KP:Approximation}.
Note that the best known result for this problem  for general graphs
is by Ravi \cite{Ra:Rapid} who provides an approximation algorithm
performance guarantee ($O(\log^2 n),O(\log n)$).

\vspace*{0.1in}
\begin{center}
\begin{tabular}{|c|c|c|c|}
  \hline

Cost Measures & Degree          & Diameter         &  Total Cost\\ \hline
Degree        &                 &                  &             \\
              & polynomial-time & polynomial-time  & polynomial-time\\ \hline 
Diameter      &                 & (weakly NP-hard) & (weakly NP-hard)\\   
              & polynomial-time & $(1 , 1 + \epsilon)$  & $(1 , 1 + \epsilon)$           \\ \hline 
Total Cost    &                 & (weakly NP-hard) & (weakly NP-hard)\\   
              & polynomial-time & $(1 , 1 + \epsilon)$    
                                             & $(1 , 1 + \epsilon)$ \\ \hline
\end{tabular}
\begin{center}
  {\bf Table 2. Bicriteria spanning tree results for treewidth-bounded
    graphs.}
\end{center}
\end{center}

\section{Hardness results}\label{sec:hardness}
The problem of finding a minimum degree spanning tree is strongly
\cclass{NP}-hard \cite{GJ:Computers}. This implies that all spanning
tree bicriteria problems, where one of the criteria is degree, are
also strongly \cclass{NP}-hard. In contrast, it is well known that the
minimum diameter spanning tree problem and the minimum cost spanning
tree problems have polynomial time algorithms
(see \cite{CLR} and the references therein).


The (Diameter, Total Cost, Spanning tree)-bicriteria problem is
strongly \cclass{NP}-hard even in the case where both cost functions
are identical \cite{HL+:Bounded}. Here we give the details of the
reduction to show that (Diameter, Total Cost, Spanning tree) is weakly
\cclass{NP}-hard even for series-parallel graphs (i.e.  graphs with
treewidth at most $2$).  Similar reductions can be
given to show that (Diameter, Diameter, Spanning tree) and (Total
cost, Total cost, Spanning tree) are also weakly \cclass{NP}-hard for
series-parallel graphs.

We first recall the definition of the {\bf PARTITION} problem
\cite{GJ:Computers}.  As an instance of the {\bf PARTITION} problem we
are given a set $T = \{ t_1,t_2, \cdots, t_n \}$ of positive integers
and the question is whether there exists a subset $X \subseteq A$ such
that 
${\displaystyle\sum_{t_i \in X} t_i = 
\sum_{t_j \in T - X} t_j  = (\sum_{t_j \in T} t_j)/2}$.

\begin{theorem} \label{np-thm} 
  (Diameter, Total cost, Spanning tree) is \cclass{NP}-hard for
  series-parallel graphs.
\end{theorem}

\begin{proof} 
  Reduction \from the {\bf PARTITION} problem. Given an instance $T =
  \{ t_1,t_2, \cdots, t_n \}$ of the {\bf PARTITION} problem, we
  construct a series parallel graph $G$ with $n+1$ vertices, $v_1,
  v_2, \cdots v_{n+1}$ and $2n$ edges. We attach a pair of parallel
  edges, $e_i^1$ and $e_i^2$, between $v_i$ and $v_{i+1}$, $1 \leq i
  \leq n$. We now specify the two cost functions $f$ and $g$ on the
  edges of this graph; $c(e_i^1) = t_i, c(e_i^2) = 0, d(e_i^1) = 0,
  d(e_i^2) = t_i, 1 \leq i \leq n$. 
Let ${\displaystyle\sum_{t_i \in T} t_i =
    2H}$. Now it is easy to show that $G$ has a spanning tree of
  $d$-diameter at most $H$ and total $c$-cost at most $H$ if and only
  if there is a solution to the original instance $T$ of the {\bf
    PARTITION} problem.

\myendofproof
\end{proof}

We now show that the (Diameter, Total-cost, Steiner tree)
problem is hard to approximate within a logarithmic factor. 
An approximation algorithm provided in Section
\ref{sec:diameter}. 
There is however a gap between the results of
Theorems \ref{fix-dia} and \ref{spanning-thm}.
Our non-approximability result is obtained by an approximation
preserving reduction from the {\bf MIN SET COVER}. 
An instance $(T,X)$ of the {\bf MIN SET COVER} problem consists of 
a universe $T = \set{t_1,t_2,\ldots,t_k}$ and a collection of subsets
$X = \{X_1, X_2,\ldots,X_m\}, ~~X_i \subseteq T$, each set $X_i$
having an associated cost $c_i$. The problem is to find a minimum cost
collection of the subsets whose union is $T$.  


\begin{fact}\label{th:feige}
Recently \cite{AS97,RS97} have independently shown the 
following non-approximability result: \\
It is $NP$-hard to find an approximate solution to the
{\bf MIN SET COVER} problem, with a universe of size $k$, with
performance guarantee better than $\Omega(\ln k)$.
\end{fact}

\begin{corollary}
\label{fix-dia}
There is an approximation preserving reduction from 
{\bf MIN SET COVER} problem to the (Diameter, Total Cost, Steiner
tree) problem. Thus:

Unless $P = NP$, given an instance of the
(Diameter, Total Cost, Steiner tree) problem with $k$ sites, there is
no polynomial-time approximation algorithm that outputs a Steiner tree
of diameter at most the bound $D$, and cost at most $R$ times that of
the minimum cost diameter-$D$ Steiner tree, for $R < \ln k$.
\end{corollary}

\begin{figure}[htbp]
  \centerline{\psfig{figure=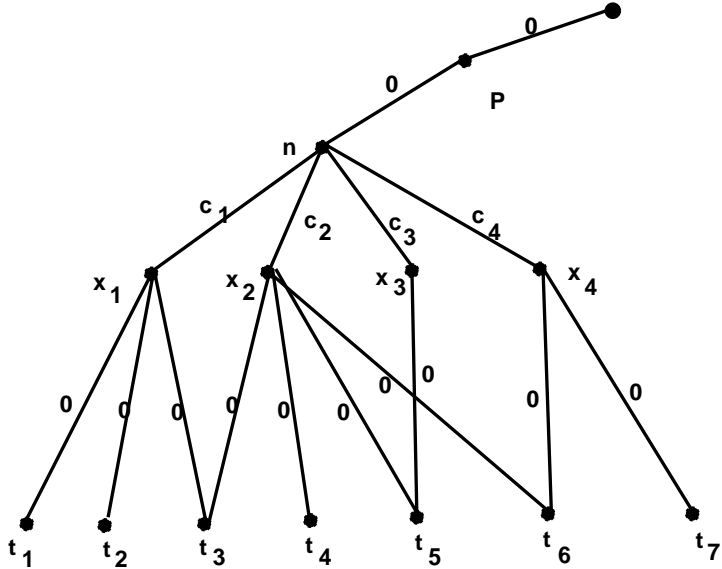,height=3in}}
\caption[hi]{Figure illustrating the reduction \from the {\bf MIN SET COVER} 
  problem to (Diameter, Total cost, Steiner tree) problem.  
The instance of {\bf MIN SET COVER} is $(T, X)$ where 
$T = \{ t_1, t_2, \ldots, t_7 \}$, $ X = \{ x_1, x_2, x_3, x_4
\}$. Here $x_1 = \{ t_1, t_2, t_3 \}, ~ x_2 = \{ t_3, t_4, t_5 \}, ~ 
x_3 = \{ t_5 \}$  and $ x_4 = \{ t_6, t_7\}$.
The cost on the edges shown in the figure denotes the $c$-cost of the edges. 
All these edges have $d$-cost $= 1$.}
\label{fig:hardapprox}
\end{figure}

\begin{proof} 
  We give an approximation preserving reduction \from the {\bf MIN SET COVER}
  problem to the (Diameter, Total Cost, Steiner tree) problem.  Given
  an instance $(T,X)$ of the {\bf MIN SET COVER} problem where $T =
  \set{t_1,t_2,\ldots,t_k}$ and $X = \{ X_1, X_2,\ldots,X_m \}, ~~ X_i
  \subseteq T$, where the cost of the set $X_i$ is $c_i$, we construct
  an instance $G$ of the (Diameter, Total Cost, Steiner tree) problem
  as follows. The graph $G$ has a node $t_i$ for each element $t_i$ of
  $T$\footnote{There is a mild abuse of notation here but it should
    not lead to any confusion.}, a node $x_i$ for each set $X_i$, and
  an extra ``enforcer-node'' $n$. For each set $X_i$, we attach an
  edge between nodes $n$ and $x_i$ of $c$-cost $c_i$, and $d$-cost
  $1$. For each element $t_i$ and set $X_j$ such that $t_i \in X_j$ we
  attach an edge $(t_i, x_j)$ of $c$-cost, 0, and $d$-cost, $1$. In
  addition to these edges, we add a path $P$ made of two edges of
  $c$-cost, 0, and $d$-cost, $1$, to the enforcer node $n$ (see Figure
  \ref{fig:hardapprox}). 
The path $P$ is added to ensure that all the nodes $t_i$ are connected
  to $n$ using a path of $d$-cost at most 2.
All other edges in the graph are assigned
  infinite $c$ and $d$-costs. The nodes $t_i$ along with $n$ and the
  two nodes of $P$ are specified to be the terminals for the Steiner
  tree problem instance. We claim that $G$ has a $c$-cost Steiner tree
  of diameter at most $4$ and cost ${\cal C}$ if and only if the
  original instance $(T,X)$ has a solution of cost ${\cal C}$.

  

Note that any Steiner tree of diameter at most $4$ must
  contain a path \from $t_i$ to $n$, for all $i$, that uses an edge
  $(x_j,n)$ for some $X_j$ such that $t_i \in X_j$. Hence any Steiner
  tree of diameter at most $4$ provides a feasible solution of
  equivalent $c$-cost to the original Set cover instance.  The proof
  now follows \from Theorem \ref{th:feige}.

\myendofproof
\end{proof}

\section{Bicriteria Formulations: Properties}
In Section~ \ref{sec:motivation}, we claimed that our formulation for
bicriteria problems is robust and general. In this section, we 
justify these claims.

\subsection{Equivalence of Bicriteria 
Formulations: Robustness}\label{sec:equivalence}

In this section, we show that our formulation for bicriteria problems
is robust and general.



Let $G$ be a graph with two (integral)\footnote{ In case of rational
cost functions, our algorithms can be extended with a small
additive loss in the performance guarantee.}
cost functions, $c$ and $d$ 
(typically edge costs or node costs). Let \cclass{A}
(\cclass{B}) be a minimization objective computed using cost function
$c$ ($d$).  Let the budget bound on the $c$-cost\footnote{We use the
  term ``cost under $c$'' or ``$c$-cost'' in this section to mean the
  value of the objective function computed using $c$, and not to mean
  the total of all the $c$ costs in the network.} ($d$-cost) of a
solution subgraph be denoted by ${\cal C}$ (${\cal D}$).

There are two natural ways to formulate a bicriteria problem: (i)
(\cclass{A}, \cclass{B}, \cclass{S}) - find a subgraph in \cclass{S}
whose \cclass{A}-objective value (under the $c$-cost) is at most ${\cal C}$
and which has minimum \cclass{B}-objective value (under the $d$-cost),
(ii) (\cclass{B}, \cclass{A}, \cclass{S}) - find a subgraph in
\cclass{S} whose \cclass{B}-objective value (under the $d$-cost) is at
most ${\cal D}$  and which has minimum \cclass{A}-objective value (under the
$c$-cost).

 
Note that bicriteria problems are generally hard, when the two
criteria are {\em hostile} with respect to each other - the
minimization of one criterion conflicts with the minimization of the
other. A good example of hostile objectives are the degree and the
total edge cost of a spanning tree in an unweighted graph
\cite{RM+:Many}. Two minimization criteria are formally defined to be
hostile whenever the minimum value of one objective is monotonically
nondecreasing as the budget (bound) on the value of the other
objective is decreased.

Let \cclass{A-APPROX}$(G,{\cal C})$ be any $(\alpha,\beta)$-approximation
algorithm for (\cclass{A}, \cclass{B}, \cclass{S}) on graph $G$ with
budget ${\cal C}$ under the $c$-cost. We now show that there is a
transformation which produces a $(\beta,\alpha)$-approximation
algorithm for (\cclass{B}, \cclass{A}, \cclass{S}). The transformation
uses binary search on the range of values of the $c$-cost with an
application of the given approximation algorithm, \cclass{A-APPROX}, at
each step of this search. Let the minimum $c$-cost of a ${\cal
  D}$-bounded subgraph in \cclass{S} be $OPT_c$. Let ${\cal C}_{hi}$
be an upper bound on the $c$-cost of any ${\cal D}$-bounded subgraph
in \cclass{S}. Note that ${\cal C}_{hi}$ is at most some polynomial in
$n$ times the maximum $c$-cost (of an edge or a node). Hence $\log
({\cal C}_{hi})$ is at most a polynomial in terms of the input
specification. Let $Heu_c$ ($Heu_d$) denote the $c$-cost ($d$-cost) of
the subgraph output by {\sc Algorithm Bicriteria-Equivalence} given
below.

\begin{center}
  \noindent \fbox{
\begin{minipage}[t]{15cm} 
  {\sc Algorithm Bicriteria-Equivalence}:
\begin{itemize}
\item {\em Input:} $G$ - graph, ${\cal D}$ - budget on criterion
  \cclass{B} under the $d$-cost, \cclass{A-APPROX} - an
  $(\alpha,\beta)$-approximation algorithm for
  (\cclass{A}, \cclass{B}, \cclass{S}).
\item
      \begin{enumerate}
      \item Let ${\cal C}_{hi}$ be an upper bound on the $c$-cost of
        any ${\cal D}$-bounded subgraph in \cclass{S}.
      \item Do binary search and find a ${\cal C}'$ in $[0,{\cal
          C}_{hi}]$ such that \label{binsearch-step}
                \begin{enumerate}
                \item \cclass{A-APPROX}$(G,{\cal C}')$ returns a subgraph
                  with $d$-cost greater than $\beta {\cal D}$, and
                  \label{step1}
                \item \cclass{A-APPROX}$(G,{\cal C}'+1)$ returns a subgraph
                  with $d$-cost at most $\beta {\cal D}$.
                  \label{step2}
                \end{enumerate}
              \item If the binary search in Step \ref{binsearch-step}
                fails to find a valid ${\cal C}'$ then output ``NO
                SOLUTION'' else output \cclass{A-APPROX}$(G,{\cal C}'+1)$.
        \end{enumerate}
      \item {\em Output:} A subgraph \from \cclass{S} such that its
        $c$-cost is at most $\alpha$ times that of the minimum
        $c$-cost ${\cal D}$-bounded subgraph and its $d$-cost is at
        most $\beta {\cal D}$.
\end{itemize} 
\end{minipage}
}
\end{center}

\begin{claim} 
  If $G$ contains a ${\cal D}$-bounded subgraph in \cclass{S} then
  {\sc Algorithm Bicriteria-Equivalence} outputs a subgraph \from
  \cclass{S} whose $c$-cost is at most $\alpha$ times that of the
  minimum $c$-cost ${\cal D}$-bounded subgraph and whose $d$-cost is
  at most $\beta {\cal D}$.
\end{claim}

\begin{proof} 
  Since \cclass{A} and \cclass{B} are hostile criteria it follows that
  the binary search in Step \ref{binsearch-step} is well defined.
  Assume that \cclass{S} contains a ${\cal D}$-bounded subgraph. Then,
  since \cclass{A-APPROX}$(G,{\cal C}_{hi})$ returns a subgraph with
  $d$-cost at most $\beta {\cal D}$, it is clear that {\sc Algorithm
    Bicriteria-Equivalence} outputs a subgraph in this case.  As a
  consequence of Step \ref{step1} and the performance guarantee of the
  approximation algorithm \cclass{A-APPROX}, we get that ${\cal C}'+1 \leq
  OPT_c$. By Step \ref{step2} we have that $Heu_d \leq \beta {\cal D}$
  and $Heu_c \leq \alpha ({\cal C}'+1) \leq \alpha OPT_c$.  Thus {\sc
    Algorithm Bicriteria-Equivalence} outputs a subgraph \from
  \cclass{S} whose $c$-cost is at most $\alpha$ times that of the
  minimum $c$-cost ${\cal D}$-bounded subgraph and whose $d$-cost is
  at most $\beta {\cal D}$.

\myendofproof
\end{proof}

Note however that in general the resulting $(\beta,\alpha)$-approximation
algorithm is, not  {\em strongly} polynomial since it
depends on the range of the $c$-costs. But it is a {\em
  polynomial-time} algorithm since its running time is linearly
dependent on $\log {\cal C}_{hi}$ the largest $c$-cost.  The above
discussion leads to the following theorem.

\begin{theorem} 
\label{equiv-thm} 
Any $(\alpha,\beta)$-approximation algorithm for
(\cclass{A}, \cclass{B}, \cclass{S}) can be transformed in polynomial
time into a $(\beta,\alpha)$-approximation algorithm for
(\cclass{B}, \cclass{A}, \cclass{S}).
\end{theorem}

\subsection{Comparing with other functional
 combinations: Generality}\label{sec:general}

Our formulation
is more general because
it subsumes the case where one wishes to minimize some functional
combination of the two criteria.  We briefly comment on this next.
For the purposes of illustration let \cclass{A} and  \cclass{B} 
be two objective functions and let us say that we wish to
minimize the sum of the two objectives \cclass{A} and  \cclass{B}.
Call this an (\cclass{A} $+$ \cclass{B}, \cclass{S}) problem.
Let \cclass{A-APPROX}$(G,{\cal C})$ be any $(\alpha,\beta)$-approximation
algorithm for (\cclass{A}, \cclass{B}, \cclass{S}) on graph $G$ with
budget ${\cal C}$ under the $c$-cost.
We show that $ \forall \epsilon > 0$, there is a polynomial time
$(1 + \epsilon) ~\max \{ \alpha,\beta \}$-approximation algorithm for the 
(\cclass{A} $+$ \cclass{B}, \cclass{S}) problem.
The transformation uses simple linear search in steps of $(1 +
\epsilon)$ over the
range of values of the $c$-cost with an
application of the given approximation algorithm, \cclass{A-APPROX}, at
each step of this search.
Let the optimum value for the (\cclass{A} $+$ \cclass{B}, \cclass{S}) problem
on  a graph $G$ be $OPT_{c + d} = (V_c + V_d)$, where $V_c$ and $V_d$ 
denote respectively the contribution of the two costs 
$c$ and $d$ for \cclass{A} and \cclass{B}. 
Let $Heu_c({\cal C})$ ($Heu_d({\cal C})$)
denote the $c$-cost ($d$-cost) of the subgraph output by 
\cclass{A-APPROX}$(G,{\cal C})$. Finally, let $Heu_{c + d}({\cal C})$ denote the 
value computed by {\sc Algorithm Convert}.                  

\begin{center}
  \noindent \fbox{
\begin{minipage}[t]{15cm} 
  {\sc Algorithm Convert}:
\begin{itemize}
\item {\em Input:} $G$ - graph, an $\epsilon > 0$,
  \cclass{A-APPROX} - an
  $(\alpha,\beta)$-approximation algorithm for
  (\cclass{A}, \cclass{B}, \cclass{S}).
\item
      \begin{enumerate}
      \item Let ${\cal C}_{hi}$ be an upper bound on the $c$-cost of
        any subgraph in \cclass{S}.
      \item
            Let $R = \lceil \log_{(1+ \epsilon)} {\cal C}_{hi} \rceil$
      \item
            For $j = 0$ to $R$ do
             \begin{enumerate}
             \item
                  $M_j = (1 + \epsilon)^j$
             \item 
                    Let $Heu_c(M_j)$, $Heu_d(M_j)$ 
                    denote the $c$-cost and  the $d$-cost of 
                    solution obtained by \cclass{A-APPROX}$(G, M_j)$. 
             \end{enumerate}
      \item
             Return the minimum over all $0 \leq j \leq R$, of
             ${\cal F}_j = Heu_c(M_j) + Heu_d(M_j)$. \label{stepa}
       \end{enumerate}
\item {\em Output:} A subgraph \from \cclass{S} such that the 
        sum of its $c$-cost  and its $d$-costs is at
        most $(1 + \epsilon) \max \{\alpha, \beta \} (OPT_{c+d})$.
\end{itemize} 
\end{minipage}
}
\end{center}

\begin{theorem}
Let \cclass{A-APPROX}$(G,{\cal C})$ be any $(\alpha,\beta)$-approximation
algorithm for (\cclass{A}, \cclass{B}, \cclass{S}) on graph $G$ with
budget ${\cal C}$ under the $c$-cost. Then, 
for all $\epsilon > 0$, there is a polynomial time
$(1+\epsilon)\max \{ \alpha,\beta \}$-approximation algorithm for the 
(\cclass{A} $+$ \cclass{B}, \cclass{S}) problem.
\end{theorem}

\noindent
{\bf Proof Sketch:}
Consider the iteration of the binary search in which the bound on the
$c$-cost is ${\cal R}$ such that $V_c \leq {\cal R} \leq  (1 +
\epsilon) V_c$. Notice that such a bound is considered as a result of
discretization of the interval  $[0,{\cal C}_{hi}]$.
Then as a  consequence 
of the performance guarantee of the 
approximation algorithm \cclass{A-APPROX}, 
we get that 
\[ Heu_c({\cal R} ) \leq \alpha {\cal R} \leq (1 + \epsilon) \alpha V_c.\] 
By Step \ref{stepa}, the performance guarantee of the algorithm  
\cclass{A-APPROX}, and the hostility of \cclass{A}  and \cclass{B}, 
we have that $ Heu_d({\cal R} ) \leq \beta V_d$. 
Thus $Heu_{c +d}({\cal R}) \leq (1 + \epsilon) \alpha V_c + \beta V_d \leq 
(1 + \epsilon) \max \{ \alpha,\beta \} (V_c + V_d)$.
Since  {\sc Algorithm Convert} outputs a subgraph \from
\cclass{S} the sum of whose  
$c$-cost and $d$-cost is minimized,  we have that
\[ \min_{{\cal C}' \in [0,{\cal C}_{hi}]}
\left( Heu_c({\cal C}') + Heu_d({\cal C}') \right) 
\leq (1 + \epsilon) \max \{ \alpha,\beta \} (OPT_{c + d}).\]
\myendofproof

A similar argument shows that 
an $(\alpha,\beta)$-approximation
algorithm \cclass{A-APPROX}$(G,{\cal C})$, 
for a (\cclass{A}, \cclass{B}, \cclass{S}) problem can
be used to find devise a polynomial time 
$(1 + \epsilon)^2 \alpha \beta$ approximation algorithm  for the
(\cclass{A} $\times$ \cclass{B}, \cclass{S}) problem.
A similar argument can also 
be given for other  basic functional combinations.
We make two additional remarks.

\begin{enumerate}

\item 
Algorithms for solving ($f$(\cclass{A}, \cclass{B}),
    \cclass{S}) problems can not in general guarantee any bounded
    performance ratios for solving the (\cclass{A}, \cclass{B},
    \cclass{S}) problem. For example, a solution for the (Total Cost +
    Total Cost , Spanning Tree) problem or the (Total Cost/Total Cost
    , Spanning Tree) problem can not be directly used to find a good
    $(\alpha,\beta)$-approximation algorithm for the (Total Cost,
    Total Cost, Spanning Tree)-bicriteria problem.

\item 
The use of approximation algorithms for (\cclass{A}, \cclass{B},
  \cclass{S})-bicriteria problems, to solve ($f$(\cclass{A},
  \cclass{B}), \cclass{S}) problems ($f$ denotes a function
  combination of the objectives) does not always yield the best
  possible solutions.  For example problems such as (Total Cost +
  Total Cost , Spanning Tree) and (Total Cost/Total Cost , Spanning
  Tree) \cite{Ch:Minimum,Me:Applying} 
 can be solved exactly in polynomial time by direct methods but
  can only be solved approximately using any algorithm for the (Total
  Cost, Total Cost , Spanning Tree)-bicriteria problem.\footnote{This
    is true since the (Total Cost, Total Cost, Spanning
    Tree)-bicriteria problem is {\bf NP}-complete and therefore unless
    {\bf P } = {\bf NP} cannot be solved in polynomial time.}

\end{enumerate}

\section{Parametric Search}\label{sec:parametric}
In this section, we present approximation algorithms for a broad class
of bicriteria problems where both the objectives in the problem are of
the same type (e.g., both are total edge costs of some network
computed using two different costs on edges, or both are diameters of
some network calculated using two different costs etc.).

As before, let $G$ be a graph with two (integral) cost functions, $c$
and $d$. Let ${\cal C}$ denote the budget on criteria \cclass{A}.  We
assume that the $c$ and $d$ cost functions are of the same kind; i.e.,
they are both costs on edges or, costs on nodes. Let
\cclass{UVW}$(G,f)$ be any $\rho$-approximation algorithm that on
input $G$ produces a solution subgraph in \cclass{S} minimizing
criterion \cclass{A}, under the single cost function $f$.  In a mild
abuse of notation, we also let \cclass{UVW}$(G,f)$ denote the
($f$-)cost of the subgraph output by \cclass{UVW}$(G,f)$ when running
on input $G$ under cost function $f$.  We use the following additional
notation in the description of the algorithm and the proof of its
performance guarantee.  Given constants $a$ and $b$
and two cost functions $f$ and $g$, defined
on edges (nodes) of a graph, $af + bg$ denotes the composite 
function that
assigns a cost $af(e) + b g(e)$ to each edge (node) in the graph.  Let
$h(\hat{{\cal D}})$ denote the cost of the subgraph, returned by
\cclass{UVW}$(G,(\frac{\hat{{\cal D}}}{{\cal C}})c + d)$ (under the
$((\frac{\hat{{\cal D}}}{{\cal C}})c + d)$-cost function).  Let the
minimum $d$-cost of a ${\cal C}$-bounded subgraph in \cclass{S} be
$OPT_d$. Let $Heu_c$ ($Heu_d$) denote the $c$-cost ($d$-cost) of the
subgraph output by {\sc Algorithm Parametric-Search} given below.

Let $\gamma > 0$ be a fixed accuracy parameter. In what follows, we
devise a $((1+\gamma),(1+\frac{1}{\gamma}))$-approximation algorithm
for (\cclass{A}, \cclass{A}, \cclass{S}), under the two cost functions
$c$ and $d$. The algorithm consists of performing a binary search with
an application of the given approximation algorithm, \cclass{UVW}, at
each step of this search.

\begin{center}
  \noindent \fbox{
\begin{minipage}[t]{15cm} 
  {\sc Algorithm Parametric-Search}:
\begin{itemize}
\item {\em Input:} $G$ - graph, ${\cal C}$ - budget on criteria
  \cclass{A} under the $c$-cost, \cclass{UVW} - a $\rho$-approximation
  algorithm that produces a solution subgraph in \cclass{S} minimizing
  criterion \cclass{A}, under a single cost function, $\gamma$ - an
  accuracy parameter.
\item \begin{enumerate}
\item Let ${\cal D}_{hi}$ be an upper bound on the $d$-cost of any
  ${\cal C}$-bounded subgraph in \cclass{S}.
\item Do binary search and find a ${\cal D}'$ in $[0, \gamma {\cal
      D}_{hi}]$ such that \label{PSbinsearch-step}
                \begin{enumerate}
                \item \cclass{UVW}$(G,(\frac{{\cal D}'}{{\cal C}})c +
                  d)$ returns a subgraph such that $\frac{h({\cal
                      D}')}{{\cal D}'} > (1+\gamma)\rho $,
                  and\label{PSstep1}
                \item \cclass{UVW}$(G,(\frac{{\cal D}'+1}{{\cal C}})c
                  + d)$ returns a subgraph such that $\frac{h({\cal
                      D}'+1)}{({\cal D}'+1)} \leq (1+\gamma)\rho $.
                  \label{PSstep2}
                \end{enumerate}
              \item If the binary search in Step \ref{binsearch-step}
                fails to find a valid ${\cal C}'$ then output ``NO
                SOLUTION'' else output \cclass{UVW}$(G,(\frac{{\cal
                    D}'+1}{{\cal C}})c + d)$.
        \end{enumerate}
      \item {\em Output:} A subgraph \from \cclass{S} such that its
        $d$-cost is at most $(1+\frac{1}{\gamma})\rho$ times that of
        the minimum $d$-cost ${\cal C}$-bounded subgraph and its
        $c$-cost is at most $(1+\gamma)\rho {\cal C}$.
\end{itemize} 
\end{minipage}
}
\end{center}

\begin{claim} \label{binsearch-justify} 
  The binary search, in Step \ref{PSbinsearch-step} of {\sc Algorithm
    Parametric-Search} is well-defined.
\end{claim}

\begin{proof}
  Since $(\frac{1}{R}$\cclass{UVW}$(G, f))$ is the same as
  \cclass{UVW}$(G,\frac{f}{R})$, we get
  that $\frac{h(\hat{{\cal D}})}{\hat{{\cal D}}}=$
  $\frac{1}{\hat{{\cal D}}}$ \cclass{UVW}$(G,(\frac{\hat{{\cal
        D}}}{{\cal C}})c + d) = $ \cclass{UVW}$(G,(\frac{1}{{\cal
      C}})c +
\frac{1}{\hat{{\cal D}}}d)$.
Hence $\frac{h(\hat{{\cal D}})}{\hat{{\cal D}}}$ is a monotone
nonincreasing function of $\hat{D}$. Thus the binary search in Step
\ref{PSbinsearch-step} of {\sc Algorithm Parametric-Search} is
well-defined.

\myendofproof
\end{proof}

\begin{claim} 
  If $G$ contains a ${\cal C}$-bounded subgraph in \cclass{S} then
  {\sc Algorithm Parametric-Search} outputs a subgraph \from \cclass{S}
  whose $d$-cost is at most $(1+\frac{1}{\gamma})\rho$ times that of
  the minimum $d$-cost ${\cal C}$-bounded subgraph and whose $c$-cost
  is at most $(1+\gamma)\rho {\cal C}$.  \end{claim}

\begin{proof} 
  By claim \ref{binsearch-justify} we have that the binary search in
  Step \ref{PSbinsearch-step} of {\sc Algorithm Parametric-Search} is
  well-defined.

  Assume that \cclass{S} contains a ${\cal C}$-bounded subgraph.
  Then, since \cclass{UVW}$(G,(\frac{\gamma {\cal D}_{hi}}{{\cal C}})c
  + d)$ returns a subgraph with cost at most $(1+\gamma)\rho {\cal
    D}_{hi}$, under the $((\frac{\gamma {\cal D}_{hi}}{ {\cal C}})c +
  d)$-cost function, it is clear that {\sc Algorithm
    Parametric-Search} outputs a subgraph in this case.

As a consequence of Step \ref{PSstep1} and the performance guarantee
of the approximation algorithm \cclass{UVW}, we get that
\[{\cal D}'+1 \leq \frac{ OPT_d}{\gamma}.\] By Step \ref{PSstep2} we
have that the subgraph output by {\sc Algorithm Parametric-Search} has
the following bounds on the $c$-costs and the $d$-costs.

\[ Heu_d \leq h({\cal D}'+1) \leq \rho( 1 + \gamma)({\cal D}' + 1)
\leq (1+ \frac{1}{\gamma}) \rho OPT_d\]
\[Heu_c \leq (\frac{{\cal C}}{{\cal D}'+1}) h({\cal D}'+1) \leq
(\frac{{\cal C}}{{\cal D}'+1}) (1+\gamma)\rho({\cal D}'+1) =
(1+\gamma)\rho {\cal C}.\]

Thus {\sc Algorithm Parametric-Search} outputs a subgraph \from
\cclass{S} whose $d$-cost is at most $(1+\frac{1}{\gamma})\rho$ times
that of the minimum $d$-cost ${\cal C}$-bounded subgraph and whose
$c$-cost is at most $(1+\gamma)\rho  {\cal C}$.

\myendofproof
\end{proof}

Note however that the resulting
$((1+\gamma)\rho,(1+\frac{1}{\gamma})\rho)$-approximation algorithm
for (\cclass{A}, \cclass{A}, \cclass{S}) may not be {\em strongly}
polynomial since it depends on the range of the $d$-costs. But it is a
{\em polynomial-time} algorithm since its running time is linearly
dependent on $\log D_{hi}$. Note that ${\cal D}_{hi}$ is at most some
polynomial in $n$ times the maximum $d$-cost (of an edge or a node).
Hence $\log ({\cal D}_{hi})$ is at most a polynomial in terms of the
input specification.


The above discussion leads to the following theorem.

\begin{theorem} \label{better-scale-thm} 
  Any $\rho$-approximation algorithm that produces a solution subgraph
  in \cclass{S} minimizing criterion \cclass{A} can be transformed
  into a $((1+\gamma)\rho,(1+\frac{1}{\gamma})\rho)$-approximation
  algorithm for (\cclass{A},\cclass{A},\cclass{S}).
\end{theorem}

The above theorem can be generalized \from the
bicriteria case to the multicriteria case (with appropriate worsening of the
performance guarantees) where all the objectives are
of the same type but with different cost functions.

\section{Diameter-Constrained Trees}\label{sec:diameter}
In this section, we describe {\sc Algorithm DCST}, our $(O(\log n),
O(\log n))$-approximation algorithm for (Diameter, Total cost, Steiner
tree) or the diameter-bounded minimum Steiner tree problem. Note that
(Diameter, Total cost, Steiner tree) includes (Diameter, Total cost,
Spanning tree) as a special case.  We first state the problem
formally: given an undirected graph $G = (V,E)$, with two cost
functions $c$ and $d$ defined on the set of edges, diameter bound $D$
and terminal set $K \subseteq V$, the (Diameter, Total cost, Steiner tree)
problem is to find a tree of
minimum $c$-cost connecting the set of terminals in $K$ with diameter
at most $D$ under the $d$-cost.

The technique underlying {\sc Algorithm DCST} is very general and has
wide applicability. Hence, we first give a brief synopsis of it. The
basic algorithm works in $(\log n)$ phases (iterations).  
Initially the solution
consists of the empty set.  During each phase of the algorithm we
execute a subroutine $\Omega$ to choose a subgraph to add to the
solution.  The subgraph chosen in each iteration is required to
possess two desirable properties.  First, it must not increase the
budget value of the solution by more than $D$; second, the solution
cost with respect to \cclass{B} must be no more than $OPT_c$, where
$OPT_c$ denotes the minimum $c$-cost of a ${\cal D}$ bounded subgraph
in {\bf S}.  Since the number of iterations of the algorithm is
$O(\log n)$ we get a $(\log n, \log n)$-approximation algorithm.  The
basic technique is fairly straightforward. The non-trivial part is to
devise the right subroutine $\Omega$ to be executed in each phase.
$\Omega$ must be chosen so as to be able to prove the required
performance guarantee of the solution.  We use the solution based
decomposition technique \cite{Ra:Rapid,RM+:Many} in the
analysis of our algorithm.  The basic idea (behind the solution based
decomposition technique) is to use the existence of an optimal
solution to prove that the subroutine $\Omega$ finds the desired
subgraph in each phase.

We now present the specifics of {\sc Algorithm DCST}.  The algorithm
maintains a set of connected subgraphs or {\em clusters} each with its
own distinguished vertex or {\em center}. Initially each terminal is
in a cluster by itself.  In each phase, clusters are merged in pairs
by adding paths between their centers. Since the number of clusters
comes down by a factor of $2$ each phase, the algorithm terminates in
$\lceil \log_2 |K| \rceil$ phases with one cluster. It outputs a
spanning tree of the final cluster as the solution.

\begin{center} 
\noindent \fbox{
\begin{minipage}[t]{15cm} 
  {\sc Algorithm Diameter-Constrained-Steiner-Tree (DCST):}
\begin{itemize}
\item {\em Input:} $G=(V,E)$ - graph with two edge cost functions, $c$
  and $d$, $D$ - a bound on the diameter under the $d$-cost,
  $K\subseteq V$ - set of terminals, $\epsilon$ - an accuracy
  parameter.
\item 
\begin{enumerate}
\item Initialize the set of clusters $\calc_1$ to contain $|K|$
  singleton sets, one for each terminal in $K$. For each cluster in
  $\calc$, define the single node in the cluster to be the center for
  the cluster. Initialize the phase count $i := 1$.
\item Repeat until there remains a single cluster in $\calc_i$
  \label{DCST-repeat}
               \begin{enumerate}
               \item Let the set of clusters $\calc_i =
                 \set{C_1\ldots,C_{k_i}}$ at the beginning of the
                 $i$'th phase (observe that $k_1 = |K|$).
               \item \label{DCST-paths} Construct a complete graph
                 $G_i$ as follows: The node set $V_i$ of $G_i$ is
                 $\set{v: v\mbox{ is the center of a cluster in }
                   \calc}$. Let path $P_{xy} $ be a
                 $(1+\epsilon)$-approximation to the minimum $c$-cost
                 diameter $D$-bounded path between centers $v_x$ and
                 $v_y$ in $G$.  Between every pair of nodes $v_x$ and
                 $v_y$ in $V_i$, include an edge $(v_x,v_y)$ in $G_i$
                 of weight equal to the $c$-cost of $P_{xy}$.
               \item Find a minimum-weight matching of largest
                 cardinality in $G_i$.\label{DCST-minwtmatching}
               \item \label{DCST-match} For each edge $e = (v_x,v_y)$
                 in the matching, merge clusters $C_x$ and $C_y$, for
                 which $v_x$ and $v_y$ were centers respectively, by
                 adding path $P_{xy}$ to form a new cluster $C_{xy}$.
                 The node (edge) set of the cluster $C_{xy}$ is
                 defined to be the union of the node (edge) sets of
                 $C_x, C_y$ and the nodes (edges) in $P_{xy}$. One of
                 $v_x$ and $v_y$ is (arbitrarily) chosen to be the
                 center $v_{xy}$ of cluster $C_{xy}$ and $C_{xy}$ is
                 added to the cluster set $\calc_{i+1}$ for the next
                 phase.
                \item $ i := i + 1.$
                \end{enumerate}
              \item Let $C'$, with center $v'$ be the single cluster
                left after Step \ref{DCST-repeat}. Output a shortest
                path tree of $C'$ rooted at $v'$ using the $d$-cost.
        \end{enumerate}
      \item {\em Output:} A Steiner tree connecting the set of
        terminals in $K$ with diameter at most $2\lceil \log_2 n\rceil
        D$ under the $d$-cost and of total $c$-cost at most
        $(1+\epsilon)\lceil \log_2 n\rceil$ times that of the minimum
        $c$-cost diameter $D$-bounded Steiner tree.
\end{itemize} 
\end{minipage}
} 
\end{center}

We make a few points about {\sc Algorithm DCST}:
\begin{enumerate}
\item The clusters formed in Step \ref{DCST-match} need not be
  disjoint.

\item All steps, except Step \ref{DCST-paths}, in algorithm DCST can
  be easily seen to have running times independent of the weights.  We
  employ Hassin's strongly polynomial \cclass{FPAS} for Step
  \ref{DCST-paths} \cite{Ha:Approximation}. Hassin's approximation
  algorithm for the $D$-bounded minimum $c$-cost path runs in time
  $O(|E|(\frac{n^2}{\epsilon} \log \frac{n}{\epsilon}))$. Thus {\sc
    Algorithm DCST} is a strongly polynomial time algorithm.

\item Instead of finding an exact minimum cost matching in Step
  \ref{DCST-minwtmatching}, we could find an approximate minimum cost
  matching \cite{GW:General}. This would reduce the running time of
  the algorithm at the cost of introducing a factor of $2$ to the
  performance guarantee.
\end{enumerate}

We now state some observations that lead to a proof of the
performance guarantee of {\sc Algorithm DCST}.  Assume, in what
follows, that $G$ contains a diameter $D$-bounded Steiner tree.
We also refer to each iteration of Step \ref{DCST-repeat} as a phase.


\begin{claim}
\label{log-lemma}
Algorithm DCST terminates in $\lceil \log_2 |K| \rceil$ phases.
\end{claim}

\begin{proof} 
  Let $k_i$ denote the number of clusters in phase $i$. Note that
  $k_{i+1} = \lceil \frac{k_i}{2} \rceil$ since we pair up the
  clusters (using a matching in Step \ref{DCST-match}). Hence we are
  left with one cluster after phase $\lceil \log_2 |K| \rceil$ and
  algorithm DCST terminates.

\myendofproof
\end{proof}

The next claim points out as clusters get merged,
the nodes within each cluster are not too far away 
(with respect to $d$-distance) from the center of the cluster.
This intuitively holds for the following  important reasons.
First, during each phase,   the graph $G_i$ has as its vertices, the
centers of the clusters in that iteration. As a result, 
we  merge the clusters by joining their centers in 
Step \ref{DCST-match}.
Second, in Step \ref{DCST-match},
for each pair of clusters $C_x$ and $C_y$ 
that are merged, we select one of their centers, $v_x$ or $v_y$ 
as the center $v_{xy}$ for the merged cluster $C_{xy}$.
This allows us to inductively maintain  two properties:
(i) the required distance of the nodes in a cluster to their centers
in an iteration $i$ is $iD$  and 
(ii) the center of a cluster at any given iteration is a terminal node.

\begin{claim} \label{claim:radius}
  Let $C \in \calc_i$ be any cluster in phase $i$ of algorithm DCST.
  Let $v$ be the center of $C$. Then any node $u$ in $C$ is reachable
  \from $v$ by a diameter-$iD$ path in $C$ under the $d$-cost.  
\end{claim}

\begin{proof} 
  Note that the existence of a diameter $D$-bounded Steiner tree
  implies that all paths added in Step \ref{DCST-match} have diameter
  at most $D$ under $d$-cost.  The proof now follows in
  a straightforward fashion by induction on $i$.

\myendofproof
\end{proof}

\begin{lemma} \label{dcostlemma} 
  Algorithm DCST outputs a Steiner tree with diameter at most $2
  \lceil \log_2 |K| \rceil\cdot D$ under the $d$-cost.
\end{lemma}

\begin{proof}
  The proof follows \from Claims \ref{log-lemma} and
  \ref{claim:radius}.

\myendofproof
\end{proof}

This completes the proof of performance guarantee with respect to the
$d$-cost. We now proceed to prove the performance guarantee with
respect to the $c$-costs. We first recall the following pairing lemma.

\begin{claim} \cite{RM+:Many} \label{pairing-claim} 
  Let $T$ be an edge-weighted tree with an even number of marked
  nodes.  Then there is a pairing $(v_1,w_1)$, $\ldots$, $(v_k,w_k)$
  of the marked nodes such that the $v_i-w_i$ paths in $T$ are
  edge-disjoint.
\end{claim}

\begin{claim} \label{cost-lemma} 
  Let $OPT$ be any minimum $c$-cost diameter-$D$ bounded Steiner tree
  and let $OPT_c$ denote its  $c$-cost. The weight of the largest
  cardinality minimum-weight matching found in Step \ref{DCST-match}
  in each phase $i$ of algorithm DCST is at most $(1+\epsilon) \cdot
  OPT_c$.
\end{claim}

\begin{proof}
Consider phase $i$ of algorithm DCST.
Note that since the centers at stage $i$ are a subset of the nodes in the
first iteration, the centers $v_i$ are  terminal nodes. Thus they belong
to $OPT$.
Mark those vertices in $OPT$ that correspond to the matched vertices,
$v_1, v_2, \ldots, v_{2\lfloor\frac{k_i}{2}\rfloor}$, of $G_i$ in
Step \ref{DCST-minwtmatching}. 
Then by Claim \ref{pairing-claim}
  there exists a pairing of the marked vertices, say
  $(v_1,v_2),\ldots, (v_{2\lfloor\frac{k_i}{2}\rfloor -1},
  v_{2\lfloor\frac{k_i}{2}\rfloor})$, and a set of edge-disjoint paths
  in OPT between these pairs. Since these paths are edge-disjoint
  their total $c$-cost is at most $OPT_c$. Further these paths have
  diameter at most $D$ under the $d$-cost. Hence the sum of the
  weights of the edges $(v_1,v_2),\ldots,
  (v_{2\lfloor\frac{k_i}{2}\rfloor -1},
  v_{2\lfloor\frac{k_i}{2}\rfloor})$ in $G_i$ , which forms a perfect
  matching on the set of matched vertices, is at most $(1+\epsilon)
  \cdot OPT_c$.  But in Step \ref{DCST-minwtmatching} of {\sc
    Algorithm DCST}, a minimum weight perfect matching 
   in the graph $G_i$ was found. Hence the weight of the matching found
  in Step \ref{DCST-match} in phase $i$ of {\sc Algorithm DCST} is at
  most $(1+\epsilon) \cdot OPT_c$.  

\myendofproof
\end{proof}

\begin{lemma} \label{ccostlemma} 
  Let $OPT$ be any minimum $c$-cost diameter-$D$ bounded Steiner tree
  and let $OPT_c$ denote its $c$-cost. {\sc Algorithm DCST} outputs a
  Steiner tree with total $c$-cost at most $(1+\epsilon) \lceil \log_2
  |K| \rceil\cdot OPT_c$.
\end{lemma}

\begin{proof} 
  \From Claim \ref{cost-lemma} we have that the $c$-cost of the set of
  paths added in Step \ref{DCST-match} of any phase is at most
  $(1+\epsilon)\cdot OPT_c$. By Claim \ref{log-lemma} there are a
  total of $\lceil \log_2 |K| \rceil$ phases and hence the Steiner
  tree output by {\sc Algorithm DCST} has total $c$-cost at most
  $(1+\epsilon) \lceil \log_2 |K| \rceil\cdot OPT_c$.

\myendofproof
\end{proof}

\From Lemmas \ref{dcostlemma} and \ref{ccostlemma} we have the following
theorem.

\begin{theorem} \label{spanning-thm} 
  There is a strongly polynomial-time algorithm that, given an
  undirected graph $G = (V,E)$, with two cost functions $c$ and $d$
  defined on the set of edges, diameter bound $D$, terminal set $K
  \subseteq V$ and a fixed $\epsilon > 0$, constructs a Steiner tree
  of $G$ of diameter at most $ 2 \lceil \log_2 |K| \rceil D$ under the
  $d$-costs and of total $c$-cost at most $(1+\epsilon) \lceil \log_2
  |K| \rceil$ times that of the minimum-$c$-cost of any Steiner tree
  with diameter at most $D$ under $d$.
\end{theorem}

\section{Treewidth-Bounded Graphs}\label{sec:treewidth}
In this section we consider the class of treewidth-bounded graphs and
give algorithms with improved time bounds and performance guarantees
for several bicriteria problems mentioned earlier.  We do this in two
steps. First we develop pseudopolynomial-time algorithms based on
dynamic programming. We then present a general method for deriving
fully polynomial-time approximation schemes (\cclass{FPAS}) \from the
pseudopolynomial-time algorithms.  We also demonstrate an 
application of the above results to the minimum broadcast time problem.

A class of treewidth-bounded graphs can be specified using a finite
number of primitive graphs and a finite collection of binary
composition rules.  We use this characterization for proving our
results.  A class of treewidth-bounded graphs $\Gamma$ is inductively
defined as follows \cite{BL+:Linear}.

\begin{enumerate}
\item The number of primitive graphs in $\Gamma$ is finite.

\item Each graph in $\Gamma$ has an ordered set of special nodes
  called {\bf terminals}.  The number of terminals in each graph is
  bounded by a constant, say $k$.

\item There is a finite collection of binary composition rules that
  operate only at terminals, either by identifying two terminals or
  adding an edge between terminals.  A composition rule also
  determines the terminals of the resulting graph, which must be a
  subset of the terminals of the two graphs being composed.
\end{enumerate}

\subsection{Exact Algorithms}\label{exact-algos}

\begin{theorem} \label{decomp-thm} 
  Every problem in Table 2 can be solved exactly in $O((n \cdot
  {\cal C})^{O(1)})$-time for any class of treewidth bounded graphs with no
  more than $k$ terminals, for fixed $k$ and a budget ${\cal C}$ on the first
  objective.
\end{theorem}

The above theorem states that there exist pseudopolynomial-time
algorithms for all the bicriteria problems \from Table 2 when
restricted to the class of treewidth-bounded graphs. The basic idea is
to employ a dynamic programming strategy. In fact, this dynamic
programming strategy (in conjunction with Theorem \ref{equiv-thm})
yields polynomial-time (not just pseudopolynomial-time) algorithms
whenever one of the criteria is the degree.  We illustrate this
strategy by presenting in some detail the algorithm for the
diameter-bounded minimum cost spanning tree problem.

\begin{theorem} \label{thm:dmsttwdth}
  For any class of treewidth-bounded graphs with no more than $k$
  terminals, there is an $O(n \cdot k^{2k+4}\cdot {\cal D}^{O(k^4)})$-time
  algorithm for solving the diameter ${\cal D}$-bounded minimum $c$-cost
  spanning tree problem.
\end{theorem}

\begin{proof} 
  Let $d$ be the cost function on the edges for the first objective
  (diameter) and $c$, the cost function for the second objective
  (total cost).  Let $\Gamma$ be any class of decomposable graphs. Let
  the maximum number of terminals associated with any graph $G$ in
  $\Gamma$ be $k$.  Following \cite{BL+:Linear}, it is assumed that a
  given graph $G$ is accompanied by a parse tree specifying how $G$ is
  constructed using the rules and that the size of the parse tree is
  linear in the number of nodes.

Let $\pi$ be a partition of the terminals of $G$. For every terminal
$i$ let $d_i$ be a number in $\{1, 2,  \ldots, {\cal D} \}$. For every pair of
terminals $i$ and $j$ in the same block of the partition $\pi$ let
$d_{ij}$ be a number in $\{1, 2,  \ldots,  {\cal D} \}$. 
Corresponding to every
partition $\pi$, set $\{d_i\}$ and set $\{d_{ij}\}$ we associate a
cost for $G$ defined as follows:\\  
\begin{tabbing}
  $Cost^{\pi}_{\{d_i\},\{d_{ij}\}} =$ \= Minimum total cost under the
  $c$ function of any forest containing\\ \> a tree for each block of
  $\pi$, such that the terminal nodes \\ \> occurring in each tree are
  exactly the members of the corresponding\\ \> block of $\pi$, no
  pair of trees is connected, every vertex in $G$\\ \> appears in
  exactly one tree, $d_i$ is an upper bound on the maximum\\ \>
  distance (under the $d$-function) \from $i$ to any vertex in the
  same\\ \> tree and $d_{ij}$ is an upper bound the distance (under
  the $d$-function)\\ \> between terminals $i$ and $j$ in their tree.
\end{tabbing}
For the above defined cost, if there is no forest satisfying the
required conditions the value of $Cost$ is defined to be $\infty$.

Note that the number of cost values associated with any graph in
$\Gamma$ is $O(k^k\cdot {\cal D}^{O(k^2)})$. We now show how the cost values
can be computed in a bottom-up manner given the parse tree for $G$. To
begin with, since $\Gamma$ is fixed, the number of primitive graphs is
finite.  For a primitive graph, each cost value can be computed in
constant time, since the number of forests to be examined is fixed.
Now consider computing the cost values for a graph $G$ constructed
\from subgraphs $G_1$ and $G_2$, where the cost values for $G_1$ and
$G_2$ have already been computed. Notice that any forest realizing a
particular cost value for $G$ decomposes into two forests, one for
$G_1$ and one for $G_2$ with some cost values. Since we have
maintained the best cost values for all possibilities for $G_1$ and
$G_2$, we can reconstruct for each partition of the terminals of $G$
the forest that has minimum cost value among all the forests for this
partition obeying the diameter constraints. We can do this in time
independent of the sizes of $G_1$ and $G_2$ because they interact only
at the terminals to form $G$, and we have maintained all relevant
information.

Hence we can generate all possible cost values for $G$ by considering
combinations of all relevant pairs of cost values for $G_1$ and $G_2$.
This takes time $O(k^4)$ per combination for a total time of
$O(k^{2k+4}\cdot {\cal D}^{O(k^4)})$. As in \cite{BL+:Linear}, we assume that
the size of the given parse tree for $G$ is $O(n)$. Thus the dynamic
programming algorithm takes time $O(n\cdot k^{2k+4}\cdot {\cal D}^{O(k^4)})$.
This completes the proof.

\myendofproof
\end{proof}

\subsection{Fully Polynomial-Time Approximation Schemes}

The pseudopolynomial-time algorithms described in the previous section
can be used to design fully polynomial-time approximation schemes
(\cclass{FPAS}) for these same problems for the class of
treewidth-bounded graphs.  We illustrate our ideas once again by
devising an \cclass{FPAS} for the (Diameter, Total cost, Spanning
tree)-bicriteria problem for the class of treewidth-bounded graphs.
The basic technique underlying our algorithm, {\sc Algorithm
  FPAS-DCST}, is approximate binary search using rounding and scaling
- a method similar to that used by Hassin \cite{Ha:Approximation} and
Warburton \cite{Wa:Approximation}.

As in the previous subsection, let $G$ be a treewidth-bounded graph
with two (integral) edge-cost functions $c$ and $d$. Let $D$ be a
bound on the diameter under the $d$-cost. Let $\epsilon$ be an
accuracy parameter. Without loss of generality we assume that
$\frac{1}{\epsilon}$ is an integer. We also assume that there exists a
$D$-bounded spanning tree in $G$. Let $OPT$ be any minimum $c$-cost
diameter $D$-bounded spanning tree and let $OPT_c$ denote its
$c$-cost.  Let \cclass{TCSTonTW}$(G,c,d,C)$ be a pseudopolynomial time
algorithm for the (Total cost, Diameter, Spanning tree) problem on
treewidth-bounded graphs; i.e., \cclass{TCSTonTW} outputs a minimum
diameter spanning tree of $G$ with total cost at most $C$ (under the
$c$-costs). Let the running time of \cclass{TCSTonTW} be $p(n,C)$ for
some polynomial $p$. For carrying out our approximate binary search we
need a testing procedure {\sc Procedure Test(V)} which we detail
below:
 
\begin{center}
  \noindent \fbox{
\begin{minipage}[t]{15cm} 
  {\sc Procedure Test($\lambda$)}:
\begin{itemize}
\item {\em Input:} $G$ - treewidth bounded graph, $D$ - bound on the
  diameter under the $d$-cost, $\lambda$ - testing parameter,
  \cclass{TCSTonTW} - a pseudopolynomial time algorithm for the (Total
  cost, Diameter, Spanning tree) problem on treewidth-bounded graphs,
  $\epsilon$ - an accuracy parameter.
\item 
  \begin{enumerate}
  \item Let $\lfloor \frac{c}{\lambda\epsilon/(n-1)} \rfloor$ denote the
    cost function obtained by setting the cost of edge $e$ to $\lfloor
    \frac{c_e}{\lambda\epsilon/(n-1)} \rfloor$.
  \item If there exists a $C$ in
    $[0,\frac{n-1}{\epsilon}]$ such that
    \cclass{TCSTonTW}$(G,\lfloor\frac{c}{\lambda\epsilon/(n-1)}\rfloor,d,C)$
    produces a spanning tree with diameter at most $D$ under the
    $d$-cost then output LOW otherwise output HIGH.
  \end{enumerate}
\item {\em Output:} HIGH/LOW.
\end{itemize} 
\end{minipage}
}
\end{center}

We now prove that {\sc Procedure Test($\lambda$)} has the properties we need
to do a binary search.

\begin{claim}\label{claim:testerhilo}
  If $OPT_c \leq \lambda$ then {\sc Procedure Test($\lambda$)} outputs LOW.  And, if
  $OPT_c > \lambda(1+\epsilon)$ then {\sc Procedure Test($\lambda$)} outputs
  HIGH.
\end{claim}

\begin{proof}
If $OPT_c  \leq \lambda$ then since 
\[{\displaystyle \sum_{e\in
    OPT}\lfloor\frac{c_e}{\lambda\epsilon/(n-1)}\rfloor \leq \sum_{e\in
    OPT}\frac{c_e}{\lambda\epsilon/(n-1)} \leq \frac{OPT_c}{\lambda\epsilon/(n-1)}
  \leq \frac{n-1}{\epsilon}  }\]
therefore {\sc Procedure Test($\lambda)$} outputs LOW.

Let $T_c$ be the $c$-cost of any diameter $D$ bounded spanning tree.
Then we have $T_c \geq OPT_c$. If $OPT_c > \lambda(1+\epsilon)$ then since
\[{\displaystyle \sum_{e\in
    T}\lfloor\frac{c_e}{\lambda\epsilon/(n-1)}\rfloor \geq \sum_{e\in
    T}(\frac{c_e}{\lambda\epsilon/(n-1)} - 1) \geq
  \frac{T_c}{\lambda\epsilon/(n-1)} - (n-1) \geq
  \frac{OPT_c}{\lambda\epsilon/(n-1)} - (n-1) >\frac{n-1}{\epsilon} }\]
therefore {\sc Procedure Test($\lambda$)} outputs HIGH.

\myendofproof
\end{proof}

\begin{claim}\label{claim:testerpoly}
  The running time of {\sc Procedure Test($\lambda$)} is
  $O(\frac{n}{\epsilon}p(n,\frac{n}{\epsilon}))$.
\end{claim}

\begin{proof}
  {\sc Procedure Test($\lambda$)} invokes \cclass{TCSTonTW} only
  $\frac{n-1}{\epsilon}$ times. And each time the budget $C$ is
  bounded by $\frac{n-1}{\epsilon}$, hence the running time of {\sc
    Procedure Test($\lambda$)} is
  $O(\frac{n}{\epsilon}p(n,\frac{n}{\epsilon}))$.

\myendofproof
\end{proof}

We are ready to describe  {\sc Algorithm FPAS-DCST} - which
uses {\sc Procedure Test($\lambda$)} to do an approximate binary search.

\begin{center}
  \noindent \fbox{
\begin{minipage}[t]{15cm} 
  {\sc Algorithm FPAS-DCST}:
\begin{itemize}
\item {\em Input:} $G$ - treewidth-bounded graph, $D$ - bound on the
  diameter under the $d$-cost, \cclass{TCSTonTW} - a pseudopolynomial
  time algorithm for the (Total cost, Diameter, Spanning tree) problem
  on treewidth-bounded graphs, $\epsilon$ - an accuracy parameter.
\item 
  \begin{enumerate}
  \item Let $C_{hi}$ be an upper bound on the $c$-cost of any
    $D$-bounded spanning tree. Let $LB = 0$ and $UB = C_{hi}$.
  \item While $UB \geq 2LB$ do \label{fpas-stepbound}
    \begin{enumerate}
    \item Let $\lambda = (LB + UB)/2$.
    \item If {\sc Procedure Test($\lambda$)} returns HIGH then set $LB = \lambda$
      else set $UB = \lambda(1+\epsilon)$.
    \end{enumerate}
  \item Run \label{fpas-stepout}
    \cclass{TCSTonTW}$(G,\lfloor\frac{c}{LB\epsilon/(n-1)}\rfloor,d,C)$
    for all $C$ in $[0,2(\frac{n-1}{\epsilon})]$ and among all the
    trees with diameter at most $D$ under the $d$-cost output the tree
    with the lowest $c$-cost.
  \end{enumerate}
\item {\em Output:} A spanning tree with diameter at most $D$ under
  the $d$-cost and with $c$-cost at most $(1+\epsilon)$ times that of
  the minimum $c$-cost $D$-bounded spanning tree.
\end{itemize} 
\end{minipage}
}
\end{center}

\begin{lemma}\label{correct}
  If $G$ contains a $D$-bounded spanning tree then {\sc Algorithm
    FPAS-DCST} outputs a spanning tree with diameter at most $D$ under
  the $d$-cost and with $c$-cost at most $(1+\epsilon)OPT_c$.
\end{lemma}

\begin{proof}
  It follows easily \from Claim \ref{claim:testerhilo} that the loop in
  Step \ref{fpas-stepbound} of {\sc Algorithm FPAS-DCST} executes
  $O(\log C_{hi})$ times before exiting with $LB \leq OPT_c \leq UB <
  2LB$.
  
  Since 
  \[{\displaystyle \sum_{e\in
      OPT}\lfloor\frac{c_e}{LB\epsilon/(n-1)}\rfloor \leq \sum_{e\in
      OPT}\frac{c_e}{LB\epsilon/(n-1)} \leq
    \frac{OPT_c}{LB\epsilon/(n-1)} \leq 2(\frac{n-1}{\epsilon}) }\] we
  get that Step \ref{fpas-stepout} of {\sc Algorithm FPAS-DCST}
  definitely outputs a spanning tree. Let $Heu$ be the tree output.
  Then we have that
  \[{\displaystyle Heu_c = \sum_{e\in Heu_c}c_e \leq
    LB\epsilon/(n-1)\sum_{e\in Heu_c}\frac{c_e}{LB\epsilon/(n-1)} \leq
    LB\epsilon/(n-1)(\sum_{e\in
      Heu_c}\lfloor\frac{c_e}{LB\epsilon/(n-1)}\rfloor + 1). }\] But
  since Step \ref{fpas-stepout} of {\sc Algorithm FPAS-DCST} outputs
  the spanning tree with minimum $c$-cost we have that
  \[{\displaystyle \sum_{e\in
      Heu_c}\lfloor\frac{c_e}{LB\epsilon/(n-1)}\rfloor \leq \sum_{e\in
      OPT}\lfloor\frac{c_e}{LB\epsilon/(n-1)}\rfloor. } \]
  Therefore 
  \[{\displaystyle Heu_c \leq LB\epsilon/(n-1)\sum_{e\in
      OPT}\lfloor\frac{c_e}{LB\epsilon/(n-1)}\rfloor + \epsilon LB
    \leq \sum_{e\in OPT}c_e + \epsilon OPT_c \leq (1+\epsilon)OPT_c.
    }\]
  This proves the claim.

\myendofproof
\end{proof}

\begin{lemma}\label{runtime}
  The running time of {\sc Algorithm FPAS-DCST} is
  $O(\frac{n}{\epsilon}p(n,\frac{n}{\epsilon})\log C_{hi})$.
\end{lemma}

\begin{proof}
  \From Claim \ref{claim:testerpoly} we see that Step~
  \ref{fpas-stepbound} of {\sc Algorithm FPAS-DCST} takes time
  $O(\frac{n}{\epsilon}p(n,\frac{n}{\epsilon})\log C_{hi})$ while Step
  \ref{fpas-stepout} takes
  time $O(\frac{2n}{\epsilon}p(n,\frac{2n}{\epsilon}))$. Hence the
  running time of {\sc Algorithm FPAS-DCST} is
  $O(\frac{n}{\epsilon}p(n,\frac{n}{\epsilon})\log C_{hi})$.

\myendofproof
\end{proof}

Lemmas  \ref{runtime} and \ref{correct} yield:

\begin{theorem}
  For the class of treewidth-bounded graphs, there is an \cclass{FPAS}
  for the (Diameter, Total cost, Spanning tree)-bicriteria problem
  with performance guarantee $(1,1+\epsilon)$.
\end{theorem}

As mentioned before, similar theorems hold for the other problems in
Table 2 and all these results extend directly to Steiner trees.

\subsection{Near-Optimal Broadcast Schemes} 

The polynomial-time
algorithm for the (Degree, Diameter, Spanning tree)-bicriteria
problem for treewidth-bounded graphs can be
used in conjunction with the ideas presented in \cite{Ra:Rapid} to
obtain near-optimal broadcast schemes for the class of
treewidth-bounded graphs. As mentioned earlier, these results
generalize and improve the results of Kortsarz and Peleg
\cite{KP:Approximation}.

Given an unweighted graph $G$ and a root $r$, a {\em broadcast scheme}
is a method for communicating a message \from $r$ to all the nodes of
$G$. We consider a telephone model in which the messages are
transmitted synchronously and at each time step, any node can either
transmit or receive a message \from at most one of its neighbors. The
{\em minimum broadcast time problem} is to compute a scheme that
completes in the minimum number of time steps. Let $OPT_r(G)$ denote
the minimum broadcast time \from root $r$ and let $OPT(G) = Max_{r\in
  G} OPT_r(G)$ denote the minimum broadcast time for the graph \from
any root. The problem of computing $OPT_r(G)$ - the {\em minimum
  rooted broadcast time problem} - and that of computing $OPT(G)$ -
the {\em minimum broadcast time problem} are both \cclass{NP}-complete
for general graphs \cite{GJ:Computers}. It is easy to see that any
approximation algorithm for the minimum rooted broadcast time problem
automatically yields an approximation algorithm for the minimum
broadcast time problem with the same performance guarantee.
We refer the readers to \cite{Ra:Rapid} for more details on this problem.
Combining our approximation algorithm
for ( Diameter, Total cost, Spanning tree)-bicriteria problem
with performance guarantee $(1,1+\epsilon)$ for the class of treewidth
bounded graphs with the observations in \cite{Ra:Rapid} yields the
following theorem.

\begin{theorem}\label{th:broadcast}
  For any class of treewidth-bounded graphs there is a polynomial-time
  $O(\frac{\log n}{ \log \log n})$-approximation algorithm for the
  minimum rooted broadcast time problem and the minimum broadcast time
  problem.
\end{theorem}

\section{Concluding Remarks}\label{sec:concluding}

We have obtained the first polynomial-time approximation algorithms
for a large class of bicriteria network design problems. 
The objective function
we considered were (i) degree, (ii) diameter and (iii) total cost.
The connectivity requirements considered were spanning trees, 
Steiner trees and (in several cases) generalized Steiner trees.
Our results were based on the following three ideas:
\begin{enumerate}
\item 
A binary search method to convert an $(\alpha,
  \beta)$-approximation algorithm for
  (\cclass{A}, \cclass{B}, \cclass{S})-bicriteria problems to a $(\beta,
  \alpha)$-approximation algorithm for
  (\cclass{B}, \cclass{A}, \cclass{S})-bicriteria problems.

\item 
A parametric search technique to devise approximation algorithms
  for (\cclass{A},\cclass{A},\cclass{S})-bicriteria problems.
We note that Theorem \ref{better-scale-thm} is very general. Given {\em any}
$\rho$-approximation algorithm for minimizing the objective \cclass{A}
in the subgraph-class \cclass{S}, Theorem \ref{better-scale-thm}
allows us to produce a $(2\rho,2\rho)$-approximation algorithm for the
(\cclass{A}, \cclass{A}, \cclass{S})-bicriteria problem. 

\item A cluster based approach for devising approximation algorithms
  for certain categories of
  (\cclass{A},\cclass{B},\cclass{S})-bicriteria problems.
\end{enumerate}

We also devised pseudopolynomial time algorithms and fully polynomial
time approximation schemes for a number of bicriteria network
design problems for the class of treewidth-bounded graphs.

\subsection*{Subsequent work}

During the time when this paper was under review, important progress has been
made in improving some of the results in this paper.
Recently, Ravi and Goemans \cite{RG:Constrained} have devised 
a $(1, 1+ \epsilon)$
approximation scheme for the (Total Cost, Total Cost, Spanning tree)
problem. Their approach does not seem to extend to devising
approximation algorithms for more general subgraphs  considered here.
In \cite{KP97},  Kortsarz and Peleg  consider the
(Diameter, Total Cost, Steiner tree) problem. 
They provide polynomial 
time approximation algorithms for this problem
with performance guarantees $(2, O(\log n))$ for constant diameter bound
$D$ and $(2 + 2\epsilon, n^{\epsilon})$ for any fixed $0 < \epsilon < 1$
for general diameter bounds.
Improving the performance guarantees for one or more of the problems 
considered here remains an interesting direction for future research.

\vspace*{.2in}

\noindent {\bf Acknowledgements:} 
We would like to thank an anonymous referee for several useful
comments and suggestions. We thank Sven Krumke (University of
W\"{u}rzberg) for 
reading the paper carefully and providing several useful comments. 
In particular, both pointed an error in the original proof of
Theorem 5.3.
We thank Professors S. Arnborg and H.~L. Bodlaender 
for pointing out to us the equivalence between treewidth
bounded graphs and decomposable graphs. 
We thank A. Ramesh for
bringing \cite{KP+:Multicast} to our attention.  
We also thank Dr. V. Kompella for making his 
other papers available to us.
Finally, we thank the referees of {\em ICALP '95} for their constructive
comments and suggestions.


\newpage

\oldspacing


\begin{thebibliography}{999999} 

\bibitem[AS97]{AS97}
S. Arora and M. Sudan, ``Improved low-degree testing and its
applications,'' 
{\em Proc. 29th Annual ACM Symposium  on Theory of Computing (STOC)}, 
485-496 (1997).


\bibitem[AB+90]{AB+:Cost} B. Awerbuch, A. Baratz, and D. Peleg,
``Cost-sensitive analysis of communication protocols,'' {\em
  Proceedings of the 9th Symposium on Principles of Distributed
  Computing (PODC)}, pp. 177-187 (1990).




\bibitem[AC+93]{AC+:Algebraic} S. Arnborg, B. Courcelle, A.
Proskurowski and D.  Seese, ``An Algebraic Theory of Graph Reductions,''
{\em Journal of the ACM (JACM)}, vol. 40:5, pp. 1134-1164 (1993).



\bibitem[AK+95]{AK+:When} A. Agrawal, P. Klein and R. Ravi, ``When
trees collide: an approximation algorithm for the generalized Steiner
problem on networks,'' {\em SIAM Journal on Computing}, vol.24, pp.
440-456 (1995).

\bibitem[AL+91]{AL+:Easy} S. Arnborg, J. Lagergren and D. Seese,
``Easy Problems for Tree-Decomposable Graphs,'' {\em Journal of
  Algorithms}, vol.  12, pp. 308-340 (1991).

\bibitem[Bo88]{Bo:Dynamic} H.L. Bodlaender, ``Dynamic programming on
graphs of bounded treewidth,'' {\em Proceedings of the 15th
  International Colloquium on Automata Language and Programming}, LNCS
vol. 317, pp.  105-118 (1988).

\bibitem[BK90]{BK90} 
A. Bookstein and S.T. Klein,
``Construction of Optimal Graphs for Bit-Vector Compression,''
{\em Proc. 13th ACM-SIGIR}, 
vol.  16, pp. 387-400 (1990).


\bibitem[BL+87]{BL+:Linear} M.W. Bern, E.L. Lawler and A.L. Wong,
``Linear -Time Computation of Optimal Subgraphs of Decomposable
Graphs,'' {\em Journal of Algorithms}, vol. 8, pp. 216-235 (1987).

\bibitem[CG82]{CG:Bounded} P. M. Camerini, and G. Galbiati, ``The
bounded path problem,'' {\em SIAM Journal on Algebraic and Discrete
  Methods} vol. 3, no. 4, pp. 474-484 (1982).


\bibitem[Ch77]{Ch:Minimum} R. Chandrasekaran,
``Minimum Ratio Spanning Trees,''
{\em  Networks,}
vol. 7, pp. 335-342, (1977).

\bibitem[CLR]{CLR} T.H. Cormen, C.E. Leiserson, and R.L. Rivest,
{\em Introduction to  Algorithms}, 
McGraw-Hill Book Co., 1990.





\bibitem[Ch91]{Ch:Multicast} C.-H. Chow, ``On multicast path finding
algorithms,'' {\em Proceedings of IEEE INFOCOM 1991}, pp. 1274-1283
(1991).


\bibitem[CK95]{CK:Compendium} P. Crescenzi and V. Kann, ``A compendium
of \cclass{NP} optimization problems,'' Manuscript, (1995).

\bibitem[FW+85]{FW+:Multicast} A. Frank, L. Wittie, and A. Bernstein,
``Multicast communication in network computers,'' {\em IEEE Software},
vol. 2, no.  3, pp. 49-61 (1985).

\bibitem[GG+95]{GG+:Multi-weighted} J. L.  Ganley, M. J.  Golin and J.
S. Salowe, ``The multi-weighted spanning tree problem,'' 
{\em Proceedings of the First Conference on Combinatorics and
Computing  (COCOON)},
Springer Verlag, LNCS pp. 141-150 (1995).



\bibitem[GJ79]{GJ:Computers} M. R. Garey and D. S. Johnson, {\em
  Computers and intractability: A guide to the theory of
  NP-completeness}, W. H.  Freeman, San Francisco (1979).

\bibitem[GW95]{GW:General} M. X. Goemans and D. P. Williamson, ``A
general approximation technique for constrained forest problems,''
{\em SIAM Journal on Computing}, 
Vol. 24, 1995, pp. 296--317.



\bibitem[Ha92]{Ha:Approximation} R. Hassin, ``Approximation schemes
for the restricted shortest path problem,'' {\em Mathematics of
  Operations Research}, vol. 17, no. 1, pp. 36-42 (1992).



\bibitem[HL+89]{HL+:Bounded} J. Ho, D.T. Lee, C.H. Chang and C.K.
Wong, ``Bounded diameter spanning tree and related problems,'' {\em
  Proceedings of the Annual ACM Symposium on Computational Geometry},
pp.  276-282 (1989).

\bibitem[Ho95]{Ho:Approximation} D. Hochbaum, {\em Approximation
  algorithms for NP-hard problems,} D.S. Hochbaum Ed., PWS Publishing
Company, Boston, MA (1995).

\bibitem[KJ83]{KJ:Routing} B. Kadaba and J. Jaffe, ``Routing to
multiple destinations in computer networks,'' {\em IEEE Transactions
  on Communications}, Vol.  COM-31, pp. 343-351 (March 1983).


\bibitem[KR+93]{KR+:Balancing} S. Khuller, B. Raghavachari, and N.
Young, ``Balancing minimum spanning and shortest path trees,'' 
{\em Algorithmica,} 
vol. 14 (4), pp. 305-321,  (1995). 

\bibitem[KP+92A]{KP+:Multicasting} V.P. Kompella, J.C. Pasquale and
G.C. Polyzos, ``Multicasting for multimedia applications,'' {\em
  Proceedings of IEEE INFOCOM 1992} (May 1992).

\bibitem[KP+93]{KP+:Multicast} V.P. Kompella, J.C. Pasquale and G.C.
Polyzos, ``Multicast routing for multimedia communication,'' {\em
  IEEE/ACM Transactions on Networking}, pp. 286-292 (1993).

\bibitem[KP92]{KP:Approximation} G. Kortsarz and D. Peleg,
``Approximation algorithms for minimum time broadcast,'' {\em SIAM
  Journal on Discrete Mathematics}, Vol. 8, No. 3, pp. 401-427 1995.



\bibitem[KP97]{KP97} G. Kortsarz and D. Peleg,
``Approximating Shallow Light Trees,'' 
{\em  Proceedings of the Eighth Annual ACM-SIAM Symposium on Discrete
  Algorithms}, pp. 103-110 (1997).

\bibitem[Me83]{Me:Applying} N. Megiddo,
``Applying parallel computation algorithms in the design of serial 
algorithms,''
{\em Journal of the ACM (JACM)}, vol.  30, pp. 852-865, (1983).

\bibitem[Ph+93]{Ph:Network} C. Phillips,
``The Network Inhibition Problem,''
{\em Proceedings of the 25th Annual ACM Symposium on the Theory of
Computing},  pp.  776-785, (1993). 

\bibitem[Ra94]{Ra:Rapid} R. Ravi, ``Rapid rumor ramification:
approximating the minimum broadcast time,'' {\em Proceedings of the
  35th Annual IEEE Foundations of Computer Science}, pp. 202-213
(1994).

\bibitem[RG95]{RG:Constrained} R. Ravi and M. Goemans, ``The
constrained spanning tree problem,'' to appear in the {\em Proceedings
  of the 5th Scandinavian Workshop on Algorithmic Theory}, 1996.


\bibitem[RM+93]{RM+:Many} R. Ravi, M. V. Marathe, S. S. Ravi, D. J.
Rosenkrantz, and H.B. Hunt III, ``Many birds with one stone:
multi-objective approximation algorithms,'' {\em Proceedings of the
  25th Annual ACM Symposium on the Theory of Computing}, pp.  438-447
(1993). (Expanded version appears as Brown University Technical Report
TR-CS-92-58.)


\bibitem[RS97]{RS97}
R. Raz and S. Safra, ``A sub-constant error-probability low-degree
test, and a sub-constant error-probability PCP characterization of
NP,''  {\em Proc. 29th Annual ACM Symposium on Theory of
Computing}, 475-484 (1997).

\bibitem[Wa87]{Wa:Approximation} A. Warburton, ``Approximation of
Pareto optima in multiple-objective, shortest path problems,'' {\em
  Operations Research}, vol. 35, pp.  70-79 (1987).



\bibitem[ZP+94]{ZP+:Iterative} Q. Zhu, M. Parsa, and W.W.M. Dai, ``An
iterative approach for delay-bounded minimum Steiner tree
construction,'' {\em Technical Report UCSC-CRL-94-39}, UC Santa Cruz
(1994).  

\end{thebibliography}
\end{document}